\documentclass[copyright,creativecommons]{eptcs}
\providecommand{\event}{FBTC 2010} % Name of the event you are submitting to
\usepackage{graphicx}
\usepackage{url}
\usepackage{subfigure}
\usepackage{verbatim}
\usepackage{amsmath,amssymb}
\usepackage{txfonts,empheq,setspace}
\usepackage{chemarrow}

\usepackage{biopepa,prism}

\renewcommand{\sync}[1]{\syncs{#1}}

\title{Complementary approaches to understanding \\the plant circadian clock}
\author{
Ozgur E. Akman${}^{1,*}$\quad\quad Maria Luisa Guerriero${}^{1}$
\quad\quad Laurence Loewe${}^{1}$ \quad\quad Carl Troein${}^{1,2}$ \\
\institute{${}^{1}$ Centre for Systems Biology at Edinburgh,
University of Edinburgh, UK\\
${}^{2}$ School of Biological Sciences, University of Edinburgh, UK\\
${}^{*}$ Current address: School of Engineering, Computing \& Mathematics, University of Exeter, UK}}

\def\titlerunning{Complementary approaches to understanding the plant circadian clock}
\def\authorrunning{Akman, Guerriero, Loewe and Troein}

\begin{document}
\maketitle

\begin{abstract}
Circadian clocks are oscillatory genetic networks that help
organisms adapt to the 24-hour day/night cycle.
%MLG: following paragraph integrated with second paragraph of introduction
%To increase our
%mechanistic understanding of these clocks it is desirable to
%investigate simple systems that possess functional similarity with
%more complex networks.
%
The clock of  the green alga {\it Ostreococcus tauri} is
the simplest plant clock discovered so far. Its many advantages as
an experimental system facilitate the testing of
computational predictions.

We present a model of the {\it Ostreococcus} clock in the stochastic
process algebra \mbox{Bio-PEPA} and exploit its mapping to different
analysis techniques, such as ordinary differential equations,
stochastic simulation algorithms and model-checking.
The small number of molecules reported for this system tests the
limits of the continuous approximation underlying differential
equations. We investigate the difference between
continuous-deterministic and discrete-stochastic approaches.
Stochastic simulation and model-checking allow us to formulate new
hypotheses on the system behaviour, such as the presence of
self-sustained oscillations in single cells under constant light conditions.

We investigate how to model the timing of dawn and dusk in the
context of model-checking, which we use to compute how the
probability distributions of key biochemical species change over
time. These show that the relative variation in expression level is
smallest at the time of peak expression, making peak time an optimal
experimental phase marker.
Building on these analyses, we use approaches from evolutionary
systems biology to investigate how changes in the rate of mRNA degradation
impacts the phase of a key protein likely to affect fitness. We
explore how robust this circadian clock is towards such potential
mutational changes in its underlying biochemistry.
Our work shows that multiple approaches lead to a more
complete understanding of the clock.

\end{abstract}

\section{Introduction}
\label{sec:introduction}

The daily cycles in sunlight, temperature and other environmental
parameters are highly important to most organisms. To follow and
anticipate these cycles, living cells generate biochemical rhythms with
a period of approximately 24 hours (\textit{circadian}). The
majority of the known circadian clocks, including those in
eukaryotes, are based on one or more interlocking transcriptional feedback loops
between a set of key genes. Crucial to the function of the clock is its
ability to \textit{entrain} to environmental signals (i.e. to adjust its
internal rhythm by synchronising with external cycles),
so that the phase of gene expression is maintained under changes to the
length of the day (photoperiod).
Such entrainment acts through various photoreceptor
pathways, where light affects kinetic parameters of the core clock.
In addition to circadian entrainment,
a defining feature of circadian clocks is that they exhibit
continued oscillations in constant light conditions~\cite{dunlap04}.

The circadian clocks of many organisms are organised around complex
feedback loop architectures, making the determination of design principles a challenging
computational problem. Although research has revealed much about the clock of the
foremost model plant organism, \textit{Arabidopsis thaliana}, there
are still unidentified components and inconsistencies between
computational models and experimental observations~\cite{jones09}.
For this reason, to increase our mechanistic understanding of
circadian clocks, it is desirable to investigate simpler systems
that possess functional similarity with more complex networks. The
circadian clock of the green alga \textit{Ostreococcus
tauri}~\cite{corellouEtAl09} is the simplest plant clock discovered
so far, and is thus an ideal model system for understanding plant circadian
function with the help of experiments, simulations
and theory. A quantitative model describing the biochemical
reactions of the \textit{Ostreococcus} clock can serve as a focal
point for this research, yielding a low-dimensional test system
for various mathematical analysis techniques.

Bio-PEPA~\cite{ciocchetta-hillston09} is a stochastic process
algebra specifically defined to model and analyse biochemical
systems. Exploiting the defined formal mappings of Bio-PEPA models
into a number of equivalent representations, it is possible to
analyse Bio-PEPA models using different mathematical and
computational techniques, including ordinary differential equations
(ODEs), stochastic simulation algorithms (SSAs) and model-checking.

%MLG: this paragraph replaces the last sentence of previous paragraph and first sentence of next paragraph
In previous work we used both ODEs  and SSAs to model the clock of
the fungus \textit{Neurospora crassa}, demonstrating that combining
different analysis methods is important for fully quantifying
%yields a more complete understanding of
the relationship between feedback architecture and circadian
behaviour~\cite{akmanEtAl09_neurospora_clock_biopepa}. Here we build
on this approach, applying a broader range of computational
techniques to the \textit{Ostreococcus} clock.  We develop and
analyse a Bio-PEPA model of the clock, focusing on various
stochastic methods which are the most appropriate in this case due
to the low copy numbers characteristic of the system. In particular,
we exploit the automatic generation of PRISM models from Bio-PEPA to
%carry out the first application of the PRISM
%model-checker~\cite{hintonEtAl06} to a circadian model, enabling
%time-dependent probability distributions to be computed for the
%MLG: slightly changed the previous sentence as follows:
carry out a novel application of the PRISM
model-checker~\cite{hintonEtAl06} to a circadian model, computing
time-dependent probability distributions for the clock components.
We use the model to quantify the variability and robustness of the
clock's functional behaviour with respect to the following factors:
(i) internal stochastic noise, the inevitable consequence of a
system comprising a small number of molecules; (ii) environmental
changes, such as photoperiod variations and transitions between
constant light/darkness; and (iii) mutational changes that affect
the biochemical reaction rates of our model, representing
perturbations to the system that occur on an evolutionary timescale.

The rest of our paper is structured as follows. After an overview of
Bio-PEPA in Section~\ref{sec:biopepa}, the \textit{Ostreococcus}
clock is introduced in Section~\ref{sec:model}, followed by
the description of the corresponding Bio-PEPA model. In
Section~\ref{sec:methods_results} we analyse the model using
various approaches. We first use stochastic simulation to
investigate how different light conditions affect the oscillations
of the clock. We then explore approaches for modelling light entrainment
in a continuous-time Markov chain (CTMC) before using model-checking
to compute the time-dependent probability distributions of protein levels.
This enables us to identify the phase markers that are most robust to stochastic fluctuations.
Finally we use ideas from a recently developed framework for evolutionary
systems biology~\cite{loewe09} to test how mutational changes in
mRNA degradation rate affect the phase of oscillations in comparison to the inherent
stochastic noise that is present at the individual cell level. The
full Bio-PEPA model is given in Appendix~\ref{sec:appendix:biopepa}.

%Each of these approaches has unique strengths and thus a more
%complete understanding can be obtained from a combined approach~\cite{akmanEtAl09_neurospora_clock_biopepa}.

\vspace*{0ex}
\section{An overview of Bio-PEPA}
\label{sec:biopepa}

Bio-PEPA~\cite{ciocchetta-hillston09} is a
stochastic process algebra, recently developed for the modelling and
analysis of biological systems. We give here a brief overview of the
main features of the language. For a detailed presentation of
its syntax and semantics, see~\cite{ciocchetta-hillston09}.

The main components of a Bio-PEPA system are the \emph{species
components}, describing the behaviour of each species, and the
\emph{model component}, specifying all interactions and initial amounts of species.
The syntax of Bio-PEPA components
is given by:\vspace*{-.5em}
$$\vspace*{-.5em} S ::= (\alpha, \kappa) \mbox{ \texttt{op} } S \mid S + S \mid C
\quad  \mbox{with }\texttt{op} = \reactant \mid \product \mid
\activator \mid \inhibitor \mid \modifier \qquad \quad \quad P::=P
\sync{\mathcal{I}}  P \mid S(x)$$
%% LL: Shouldnt that be S[x]  i.e. square brackets since that is used in the code as well.?
\noindent where $S$ is the \emph{species component} and $P$ is the
\emph{model component}. In the prefix term $(\alpha,\kappa) \mbox{
\texttt{op} } S$, $\kappa$ is the \emph{stoichiometry coefficient}
of species $S$ in reaction $\alpha$, and the \emph{prefix
combinator} ``\texttt{op}'' represents the role of $S$ in the
reaction. Specifically, $\reactant$ indicates a \emph{reactant},
$\product$ a \emph{product}, $\activator$ an \emph{activator},
$\inhibitor$ an \emph{inhibitor} and $\modifier$ a generic
\emph{modifier}. The notation $\alpha \mbox{ \texttt{op} }$ is a shorthand for
$(\alpha, \kappa) \mbox{ \texttt{op} } S$ when $\kappa=1$.
%
% LL: I think we can justify linking this into one paragraph. Agreed?
%
The operator ``$+$'' expresses a choice between possible actions,
and the constant $C$ is defined by an equation $C \rmdef S$.
The
process $P \sync{\mathcal{I}} Q$ denotes synchronisation between
components $P$ and $Q$; the set $\mathcal{I}$ determines the
activities on which the operands are forced to synchronise, with
$\sync{*}$ denoting a synchronisation on all common action types. In
the model component $S(x)$, the parameter $x \in \Nat$ represents
the initial number of molecules $S$ present.
In addition to species and model components, a Bio-PEPA system consists of
%context defining
kinetic rates, parameters and, if needed, locations,
events and other auxiliary information for the species.
%
%%LL: I didn't delete any content from the following, but I shortened it a bit. Please check if the biopepa website *needs* to be cited as a reference. I think mentioning it in text is enough and saves us space.
% Other than here the website is only cited twice in this paper, both times for the tools. One could argue that the following mention for the tools would be enough, but I leave it up to you to decide.
%ML: I prefer the ref, I don't think it's good practice to put web pages like this rather than as refs.

The formal representation offered by Bio-PEPA allows for different kinds of analysis through the
defined mapping into contin\-uous-deterministic and
discrete-stochastic analysis methods (see~\cite{ciocchetta-hillston09} for details).
More on Bio-PEPA can be found at~\cite{biopepa_site}, including two
software tools, the Bio-PEPA Eclipse Plug-in and the Bio-PEPA
Workbench~\cite{duguidEtAl09_biopepa_tools}. Both tools process Bio-PEPA
models automatically and either compute time-series results directly
using various SSA or ODE solvers, or generate representations that
can be used by other tools.

\vspace*{0ex}
\section{The \textit{Ostreococcus} clock}
\label{sec:model}

\begin{figure}[b]
\centering
  \vspace*{-.5ex}\includegraphics[width=0.71\textwidth]{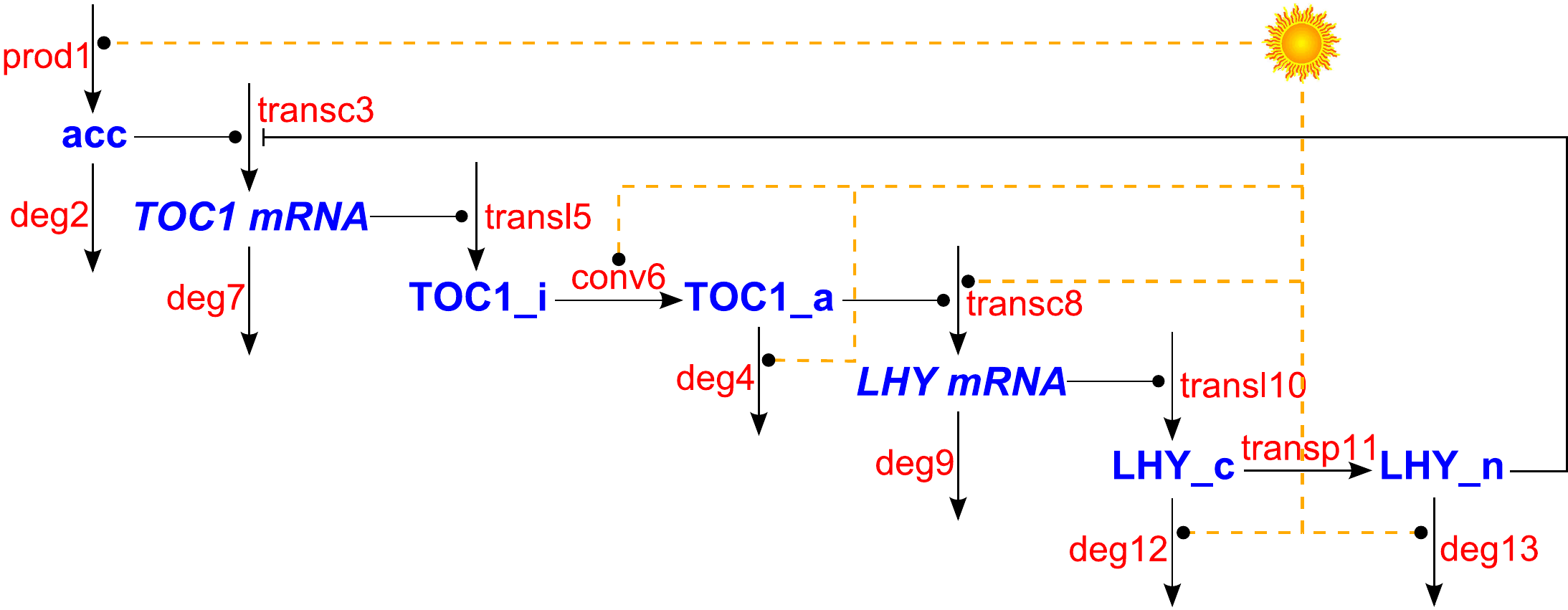}\vspace*{-1.ex}
\caption{The genetic regulatory network underlying our model of the
  \textit{Ostreococcus} clock. The network comprises a single negative
  feedback loop involving the \textit{LHY} and \textit{TOC1} genes,
  augmented by 5 light inputs which synchronise the endogenous
  oscillations in gene expression to the day/night
  cycle~\cite{oneill}.}\label{fig:model_diagram}
\end{figure}

\textit{Ostreococcus tauri} is an exceptionally small green alga with
a highly reduced genome~\cite{derelle}. Experiments and homology searches
indicate that its circadian clock is very simple compared to higher
plants, such as \textit{Arabidopsis thaliana}. Only
a handful of the clock genes identified in other plants have been found
in \textit{Ostreococcus}, and only two of these appear to be central to
the clock. The first of these, which we refer to as \textit{TOC1}, is homologous to
\textit{Arabidopsis} \textit{TOC1} (\textit{TIMING OF CAB EXPRESSION~1})
and other \textit{PRR}s (\textit{PSEUDO RESPONSE REGULATOR}s). The other gene,
here called \textit{LHY}, is homologous to \textit{Arabidopsis} \textit{LHY} (\textit{LATE
ELONGATED HYPOCOTYL}) and \textit{CCA1} (\textit{CIRCADIAN CLOCK
ASSOCIATED~1})~\cite{corellouEtAl09}.
%
%OEA: I've rewritten the above para as it was difficult to understand.
%
An ODE model of the \textit{Ostreococcus} clock as a negative feedback loop between these two genes
was introduced in~\cite{oneill}, where it was applied to drug treatments and other perturbations.
The full model includes details of the luciferase assay used to measure mRNA and protein levels,
but here we use only the central parts of the model, which describe the dynamics of the native
mRNAs and proteins. The model is illustrated in Figure~\ref{fig:model_diagram}.

%%% LL: In the following text and in the figure: Do we need to change LHY to CCA1?
%% is CCA1 also an Arabidopsis gene like LHY?
% CT: No and yes. See above.

\textit{TOC1} transcription requires light, which is buffered by a ``light accumulator'' (acc) and is
inhibited by the presence of nucleic LHY protein (LHY\_n) for most of the day. TOC1 activates \textit{LHY}
transcription through an unknown mechanism, proposed in~\cite{oneill} to work as follows: \textit{TOC1}
mRNA is translated into inactive TOC1 protein (TOC1\_i), which is activated slowly during the day but
quickly after dusk. The active form (TOC1\_a) drives \textit{LHY} transcription throughout the night
but is quickly degraded after dawn. \textit{LHY} mRNA is translated into cytosolic LHY (LHY\_c), which
is quickly translocated to the nucleus, thereby closing the feedback loop. Light also accelerates the
rate of LHY degradation.

The model parameters were estimated by fitting simulated time-courses to equivalent data
obtained from experiments over a wide range of light conditions. Some experiments alternated 12 hours of
light and dark (denoted LD 12:12), others used longer or shorter days (such as LD 16:8 or 8:16), and many
included transitions between different conditions, often into constant light (LL)~\cite{oneill}.

\subsection{A Bio-PEPA model of the clock}
\label{sec:model:biopepa}

A model of the clock as described above was
implemented in Bio-PEPA. Here we describe its main features;
for the full model, including kinetic laws and parameters, see Appendix~\ref{sec:appendix:biopepa}.
% LL: no need for paragraph here, right?
%ML: It seems we have a different taste for long vs short paragraphs. I find monolithic blocks hard to read

One of the key issues involved in obtaining a realistic stochastic model is the correct
scaling of the initial concentrations and kinetic parameters of the continuous ODE model in~\cite{oneill}
so as to obtain the respective molecule counts and rate constants for the Bio-PEPA discrete-state model.
% Note: this model based on molecular counts could be and SSA *and* ODE model!
Since the absolute values of the initial concentrations are not known, the initial values in the
original ODE model are given in arbitrary relative units. However, the peak number of TOC1 and LHY
protein molecules was estimated experimentally
over a number of free-running cycles
in LL conditions using a TopCount luminometer. From this, approximate initial values for our discrete-state
model were computed, yielding a rough estimate for the scaling factor of  $\Omega = 50$.
After such rescaling the  Bio-PEPA model can be analysed by ODEs and SSAs, which both give results
in molecule counts.

The proteins and mRNAs shown in Figure~\ref{fig:model_diagram} are
modelled as the following Bio-PEPA species components that describe
the possible reactions they can participate in and how their amounts
are affected by the occurrence of each reaction. Reactions are
associated with functional rates representing the corresponding
kinetic law.\\
\begin{tabular}[l]{l l l    l l l}
\textit{TOC1\_mRNA} \!\!\!\!\!\! & $\rmdef$ \!\!\!\!\!\! & $\mathit{transc}_3 \, \product \, + \, \mathit{transl}_5 \, \activator \, + \, \mathit{deg}_7 \, \reactant$ & \quad \textit{LHY\_mRNA} \!\!\!\!\!\! & $\rmdef$ \!\!\!\!\!\! & $\mathit{transc}_8 \, \product \, + \, \mathit{deg}_9 \, \reactant \, + \, \mathit{transl}_{10} \, \activator$\\
\textit{TOC1\_i} \!\!\!\!\!\! & $\rmdef$ \!\!\!\!\!\! & $\mathit{transl}_5 \, \product \, + \, \mathit{conv}_6 \, \reactant$ & \quad \textit{LHY\_c} \!\!\!\!\!\! & $\rmdef$ \!\!\!\!\!\! & $\mathit{transl}_{10} \, \product \, + \, \mathit{transp}_{11} \, \reactant \, + \, \mathit{deg}_{12} \, \reactant$\\
\textit{TOC1\_a} \!\!\!\!\!\! & $\rmdef$ \!\!\!\!\!\! & $\mathit{deg}_4 \, \reactant \, + \, \mathit{conv}_6 \, \product \, + \, \mathit{transc}_8 \, \activator$  & \quad\textit{LHY\_n} \!\!\!\!\!\! & $\rmdef$ \!\!\!\!\!\! &  $\mathit{transc}_3 \, \inhibitor \, + \, \mathit{transp}_{11} \, \product \, + \, \mathit{deg}_{13} \, \reactant$\\
\textit{acc} \!\!\!\!\!\! & $\rmdef$ \!\!\!\!\!\! & $\mathit{prod}_1 \, \product \, + \, \mathit{deg}_2 \, \reactant \, + \, \mathit{transc}_3 \, \activator$ & & \\[1ex]
\end{tabular}

For instance, the transcription of \textit{TOC1\_mRNA} is modelled
by reaction $\mathit{transc}_3$, which involves three different
species (\textit{TOC1\_mRNA}, \textit{LHY\_n}, and \textit{acc}),
and is positively regulated by the light-accumulator \textit{acc}
and negatively regulated by \textit{LHY\_n}. The kinetic law for
this reaction is given by a Hill function, commonly used for
describing transcription in clock
models~\cite{akmanEtAl09_neurospora_clock_biopepa,Akman08,Gonze02,loewe-hillston08}:
\vspace*{-.1ex}
$$\vspace*{-.1ex}\Omega \cdot \frac{\mathit{tmp\_toc1\_transcription}}{1 + \mathit{tmp\_toc1\_transcription} + \left(\frac{\mathit{R\_toc1\_lhy}}{\Omega} \cdot \mathit{LHY\_n}\right)^{\mathit{H\_toc1\_lhy}}} \enspace .$$
Here, species names represent molecule counts, $\mathit{tmp\_toc1\_transcription} = \mathit{L\_toc1} +  \mathit{acc} \cdot \mathit{R\_toc1\_acc} / \Omega$ and $\mathit{L\_toc1}$, $\mathit{R\_toc1\_acc}$, $\mathit{R\_toc1\_lhy}$ and $\mathit{H\_toc1\_lhy}$ are parameters.
%
%To compare our results
For comparisons with experiments we also defined the observables
$\mathit{Total\_LHY} = \mathit{LHY\_c} + \mathit{LHY\_n}$ and
$\mathit{Total\_TOC1} = \mathit{TOC1\_i} + \mathit{TOC1\_a}$.
% LL: I renamed them from total_TOC1 to TOC1_total, since TOC1 is more important than "total".  I hope that's OK (I will change this throughout the ms).
% ML: While I agree in principle, I changed them back. We had agreed on names long ago, I don't want to redo the plots again.

%
%
Bio-PEPA functional rates allow  the definition of general kinetic laws.
We use this facility to represent the
entrainment of the system to light/dark cycles through the time-dependent
function below:
\vspace*{-.1ex}
$$\vspace*{-.1ex}\mathit{light\_time} = H\left( \left(\left(\mathit{time} - 24 \cdot \left\lfloor \frac{\mathit{time}}{24} \right\rfloor\right) - t_{\mathit{dawn}}\right) \cdot \left(t_{\mathit{dusk}} - \left(\mathit{time} - 24 \cdot \left\lfloor \frac{\mathit{time}}{24} \right\rfloor\right)\right)\right) \enspace .$$
\noindent This allows us to model light-dependent reaction rates by returning
the value 1 in day-time and 0 during night-time. The parameters
$t_{\mathit{dawn}}$ and $t_{\mathit{dusk}}$ give the time of
the day (in hours) at which dawn and dusk occur, respectively;
$H(x)$ is the Heaviside step function that returns 1 for $x>0$ and 0 otherwise.

%\footnote{$H$ is the Heaviside step function
%defined as $H(x) = \left\{
%  \begin{array}{ll}
%    0 & \hbox{if $x <= 0$} \\
%    1 & \hbox{otherwise} \enspace .
%  \end{array}
%\right.$}.

\section{Analysis methods and results}
\label{sec:methods_results}

Each technique that can be used to analyse Bio-PEPA
models has its particular strengths: ordinary differential equations
(ODEs) easily predict mean values and quantify dynamical changes 
in terms of bifurcations, stochastic simulation algorithms
(SSAs) allow variability in the system's responses to be measured, and 
model-checking enables complex queries about the model to be formulated 
and verified automatically. Here we analyse the clock model using these three
analysis methods. After briefly describing each method, we explain why it is 
better suited for investigating a particular aspect of the system, and report 
some of the results obtained.

\subsection{Stochastic simulation: population versus single cell behaviour}
\label{sec:methods_results:ssa}

Following the formulation of Gillespie's stochastic simulation
algorithm~\cite{gillespie77}, the stochastic analysis of biochemical systems has received increasing attention
due to the impact that stochastic variability can have on system behaviour.
This is particularly relevant for systems such as gene regulatory
networks, where some molecules are present in copy numbers so small that
random fluctuations are too large for the continuous approximation behind
ODEs to be justified.
Within this framework, a single molecule-by-molecule stochastic
simulation run can be viewed as a faithful representation of
behaviour at the cellular level (assuming the underlying model is
accurate). Observing the mean behaviour over a larger number of runs
is then equivalent to observing a population of cells. Most current
experimental techniques only allow population-level assays. However, as progress in high-resolution imaging
techniques reduces the minimum population size that can be measured,
it will also become possible to consider the effect of stochastic noise, which is expected to be more evident in smaller populations.
%
%However,
%progress in developing high-resolution imaging techniques will soon enable measurements at
%the single cell level to be using in modelling studies, in addition to the average responses integrated over large populations of cells.

%\subsubsection{Application to the \textit{Ostreococcus} clock}
%\paragraph{}

\begin{figure}[hbt]
\centering
\subfigure[DD -- ODE]{\includegraphics[width=0.31\textwidth]{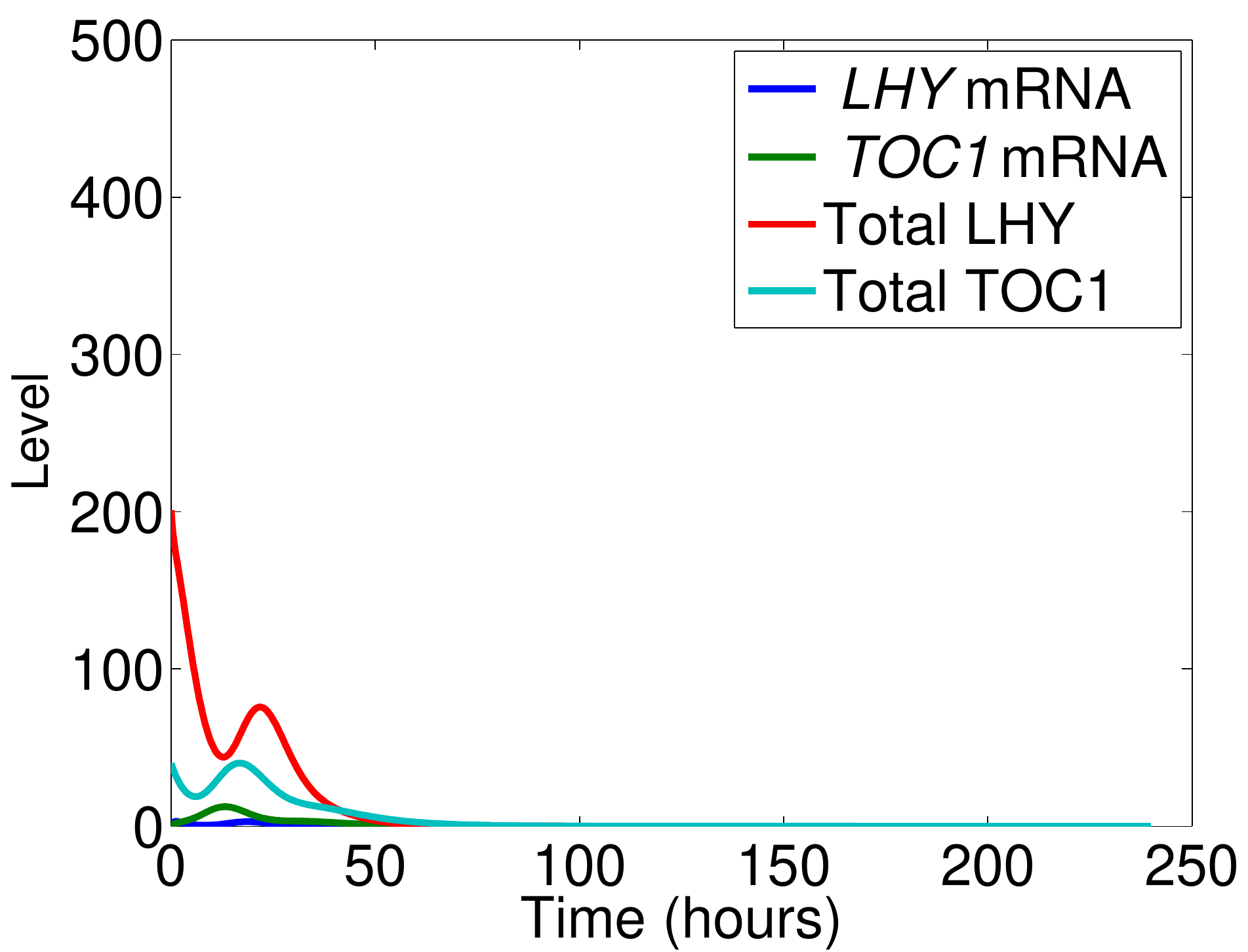}\label{fig:DD_LL_LLDD:DD_ODE}}
\subfigure[DD -- average 10000 SSA runs]{\includegraphics[width=0.31\textwidth]{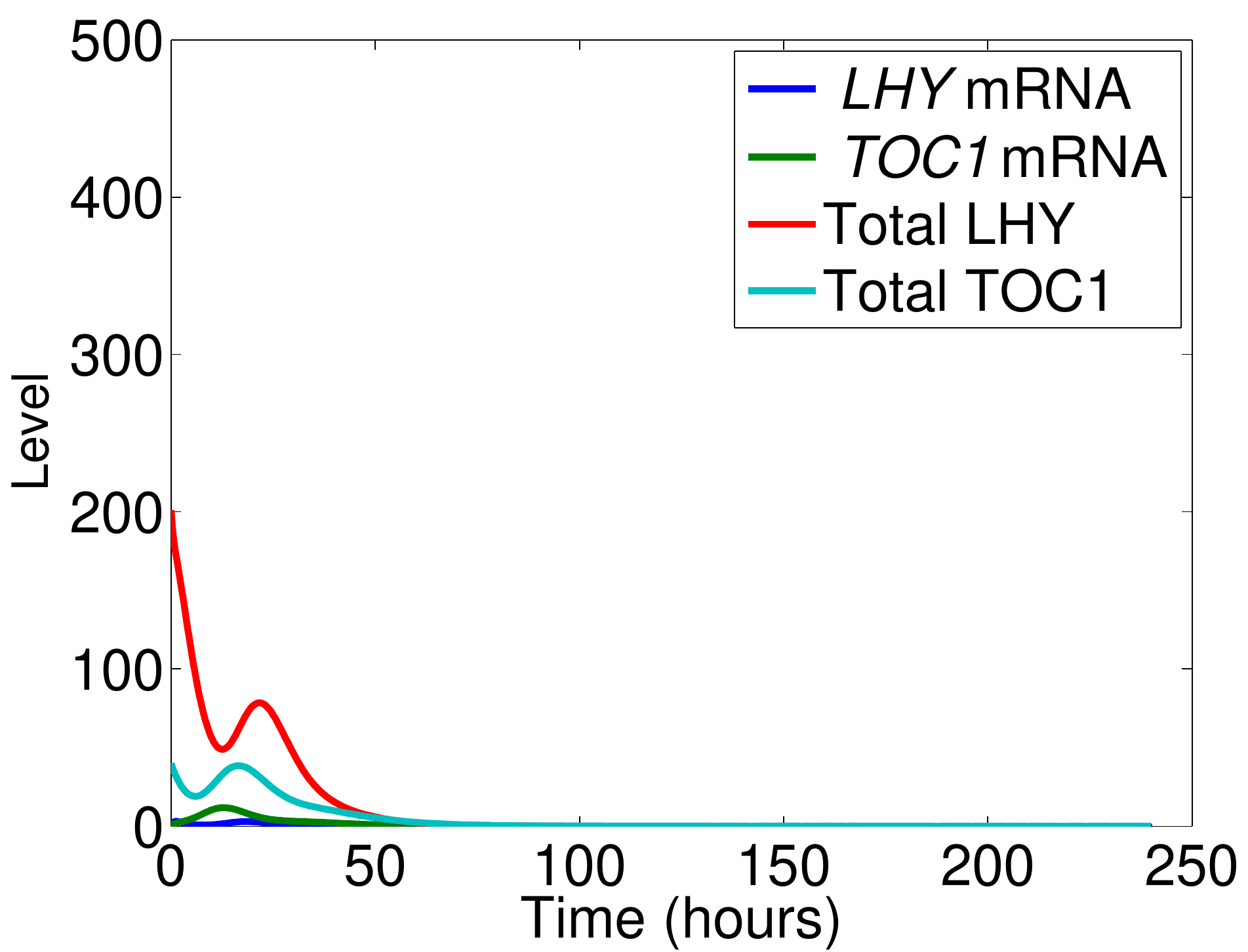}\label{fig:DD_LL_LLDD:DD_SSA10000}}
\subfigure[DD -- single SSA run]{\includegraphics[width=0.31\textwidth]{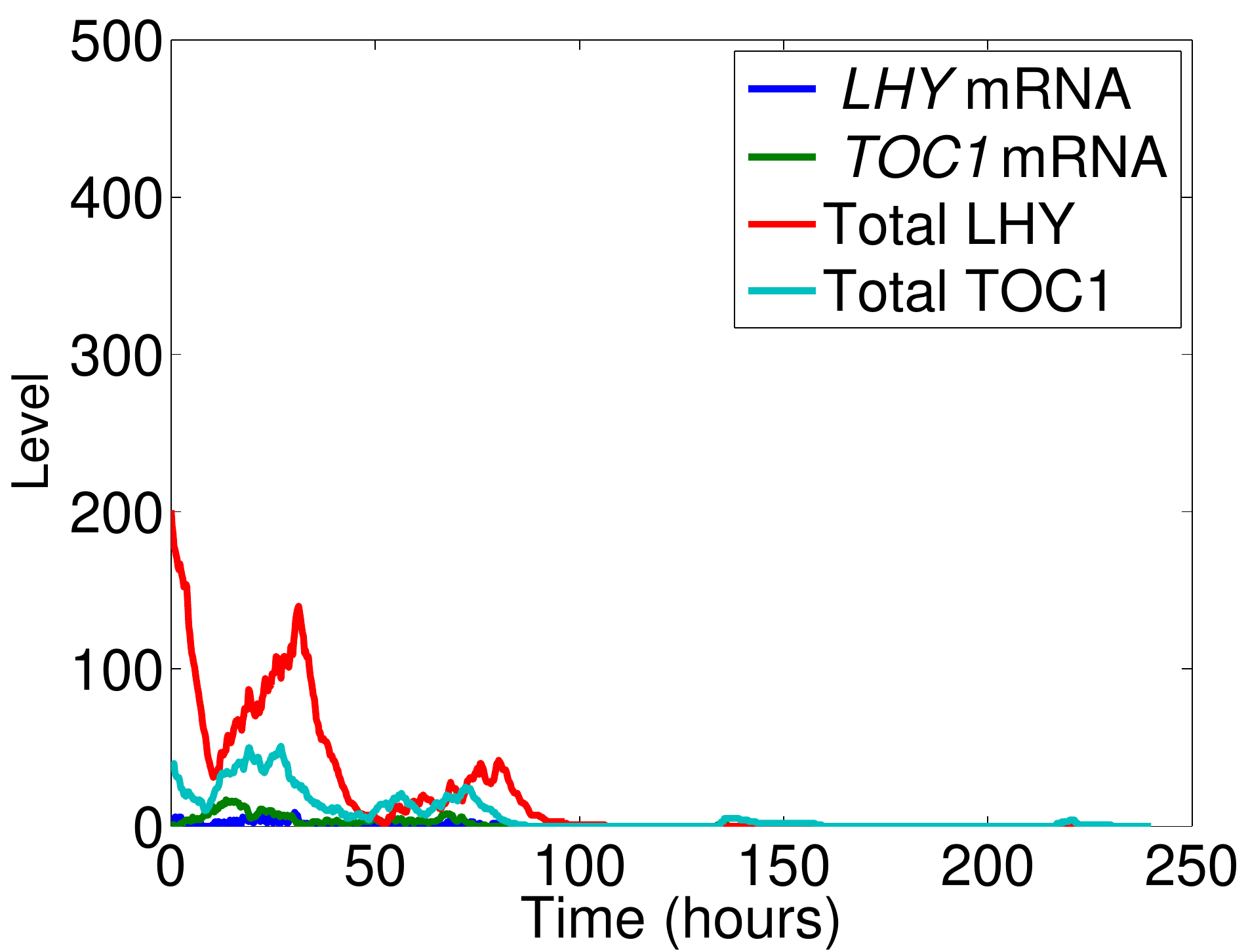}\label{fig:DD_LL_LLDD:DD_SSA1}}\\\vspace{-1.1ex}
\subfigure[LL -- ODE]{\includegraphics[width=0.31\textwidth]{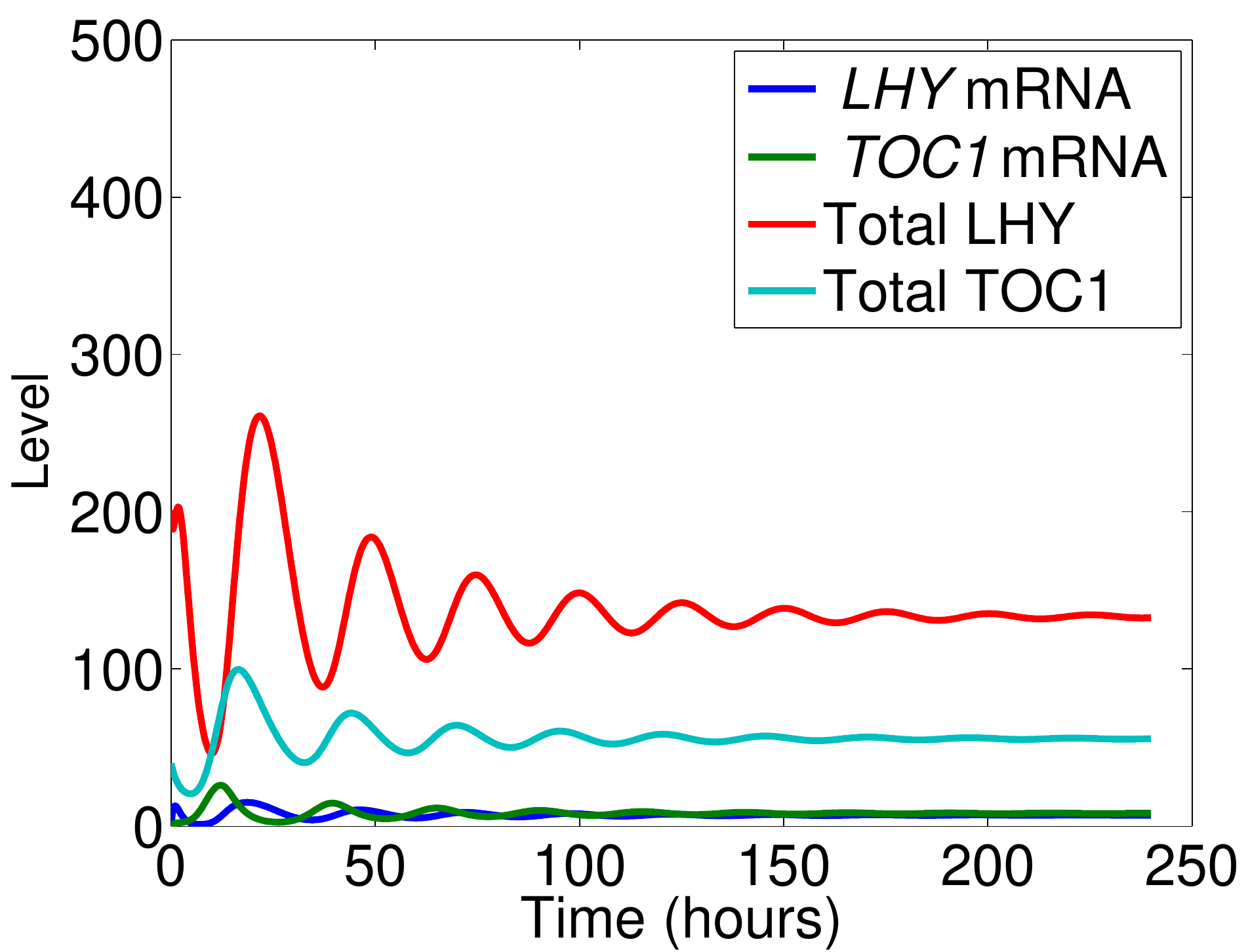}\label{fig:DD_LL_LLDD:LL_ODE}}
\subfigure[LL -- average 10000 SSA runs]{\includegraphics[width=0.31\textwidth]{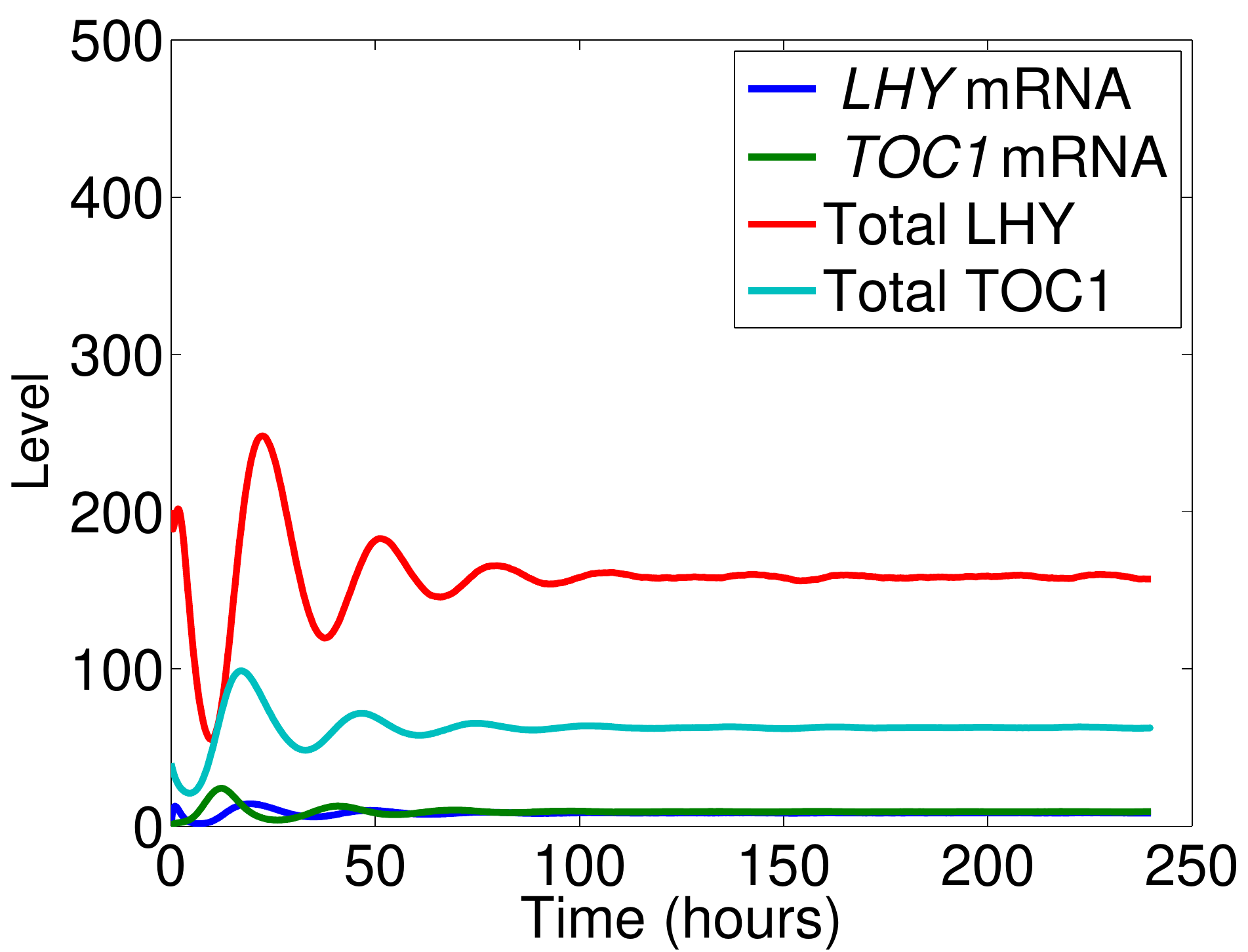}\label{fig:DD_LL_LLDD:LL_SSA10000}}
\subfigure[LL -- single SSA run]{\includegraphics[width=0.31\textwidth]{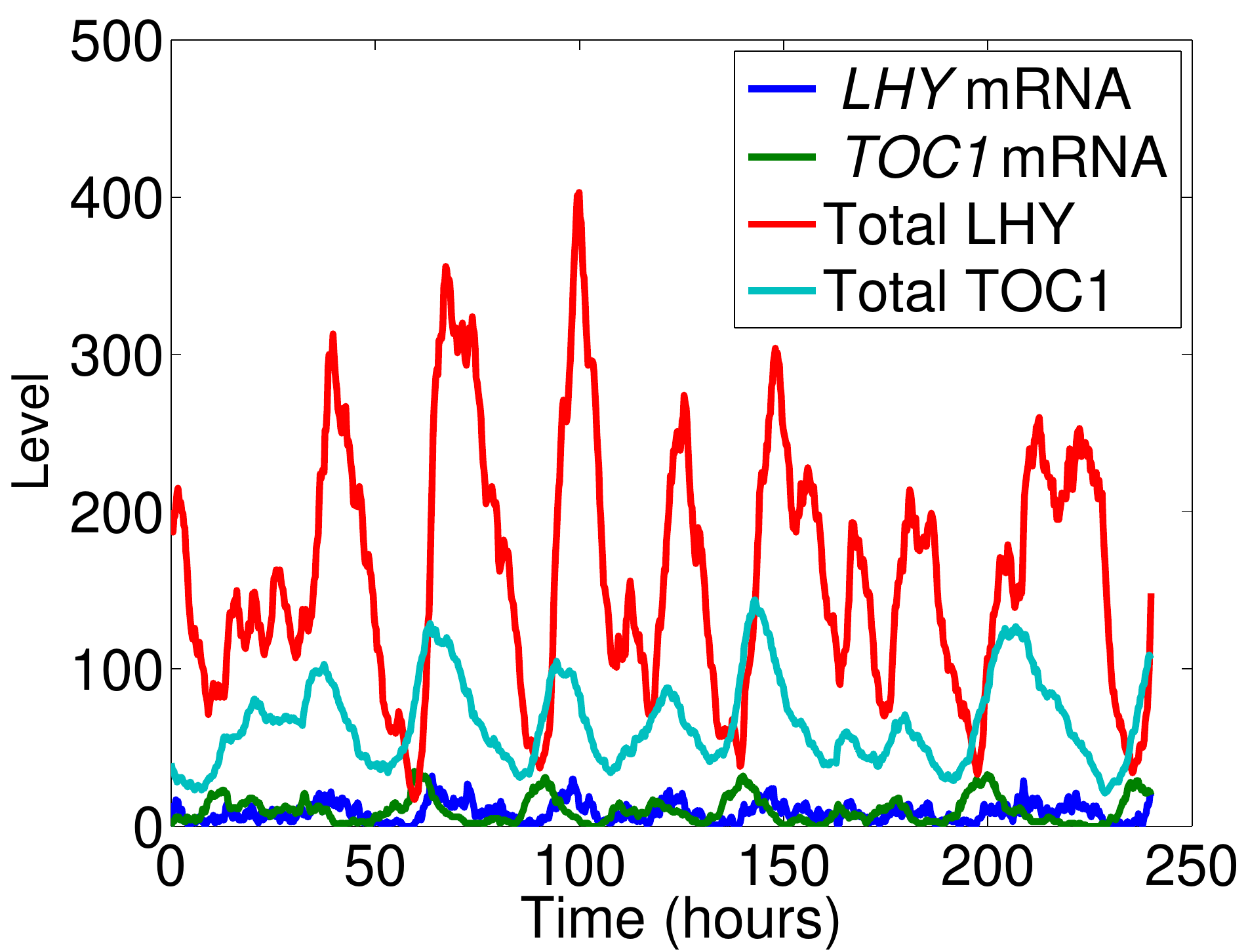}\label{fig:DD_LL_LLDD:LL_SSA1}}\\\vspace{-1.1ex}
\subfigure[LL-DD -- ODE]{\includegraphics[width=0.31\textwidth]{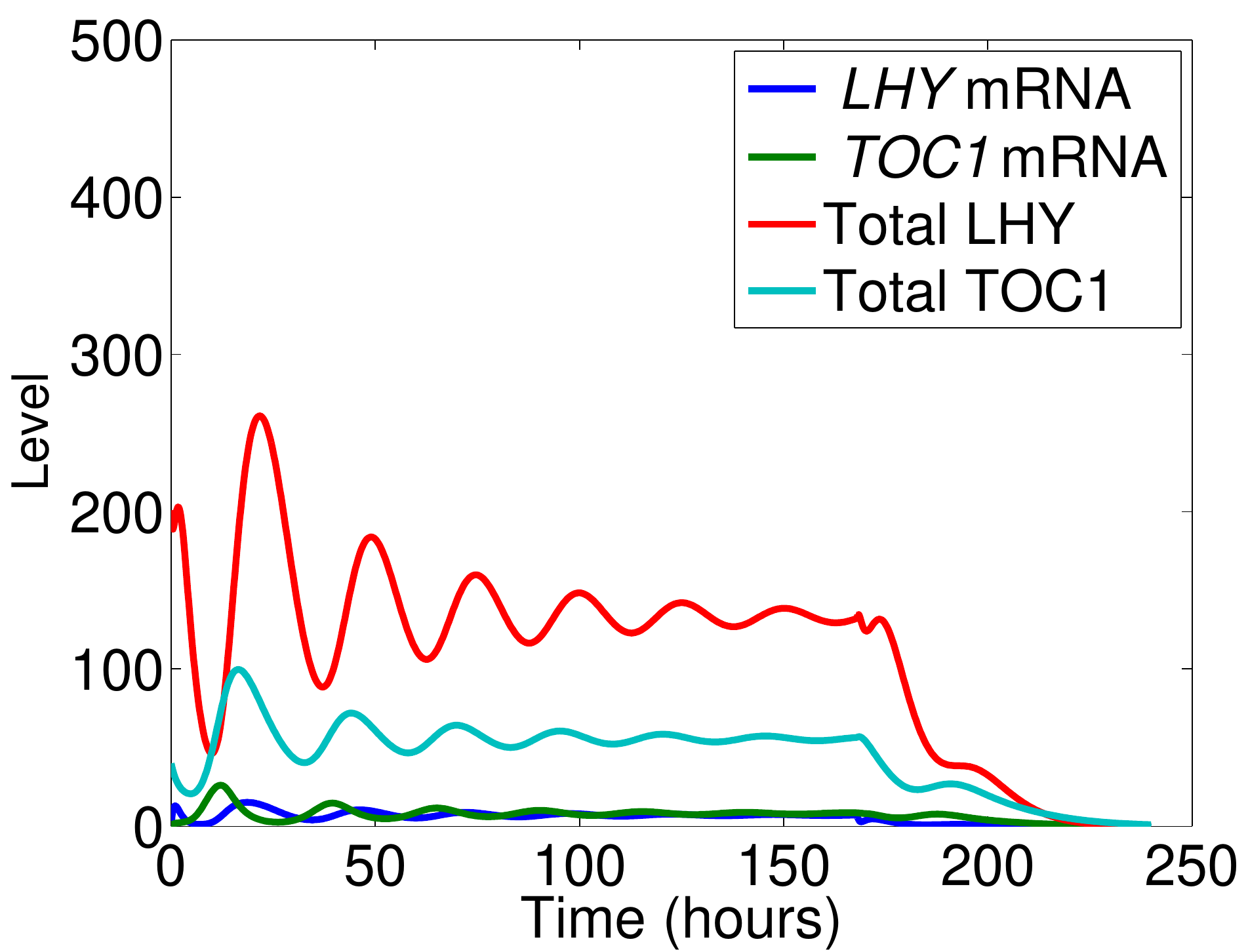}\label{fig:DD_LL_LLDD:LLDD_ODE}}
\subfigure[LL-DD -- average 10000 SSA runs]{\includegraphics[width=0.31\textwidth]{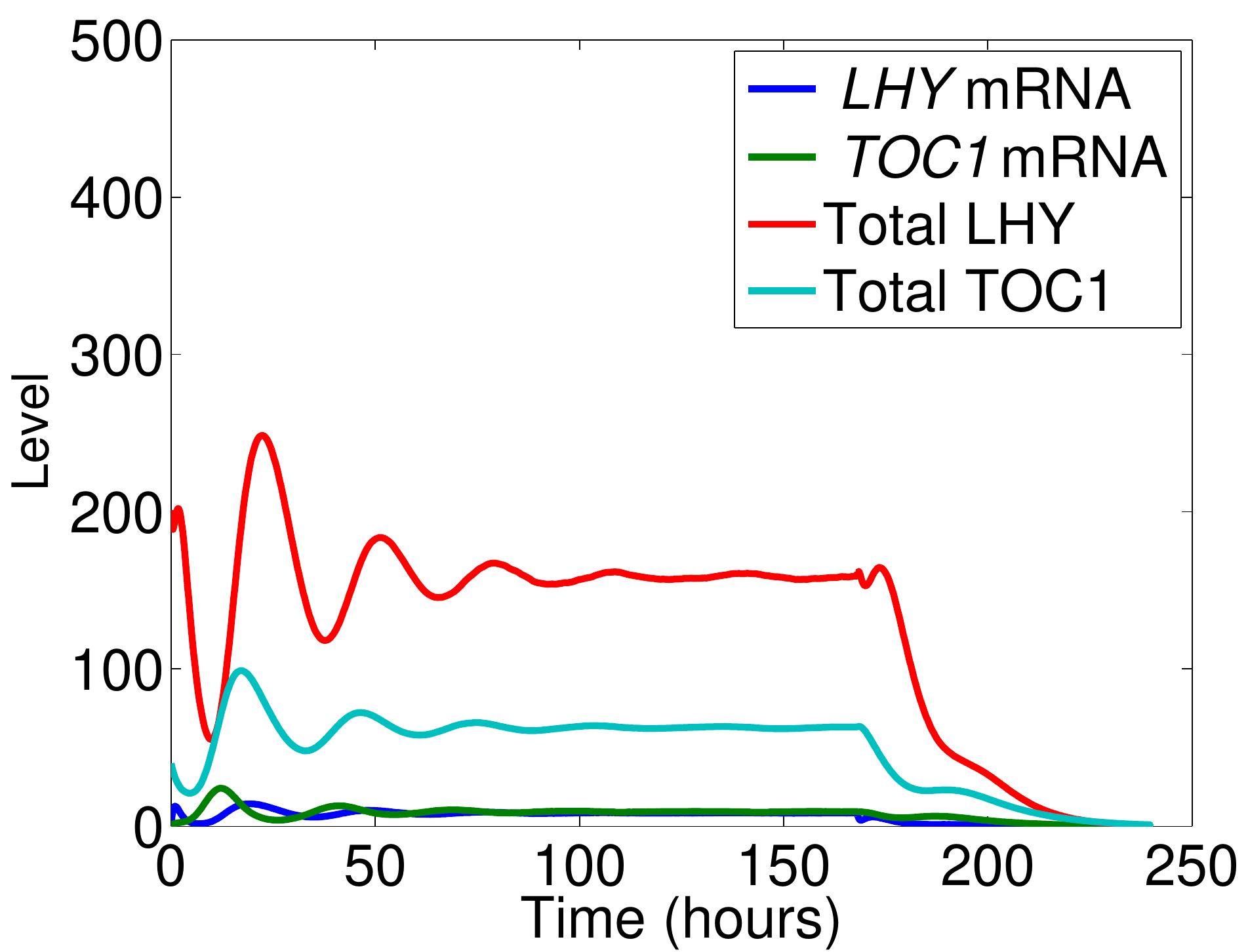}\label{fig:DD_LL_LLDD:LLDD_SSA10000}}
\subfigure[LL-DD -- single SSA run]{\includegraphics[width=0.31\textwidth]{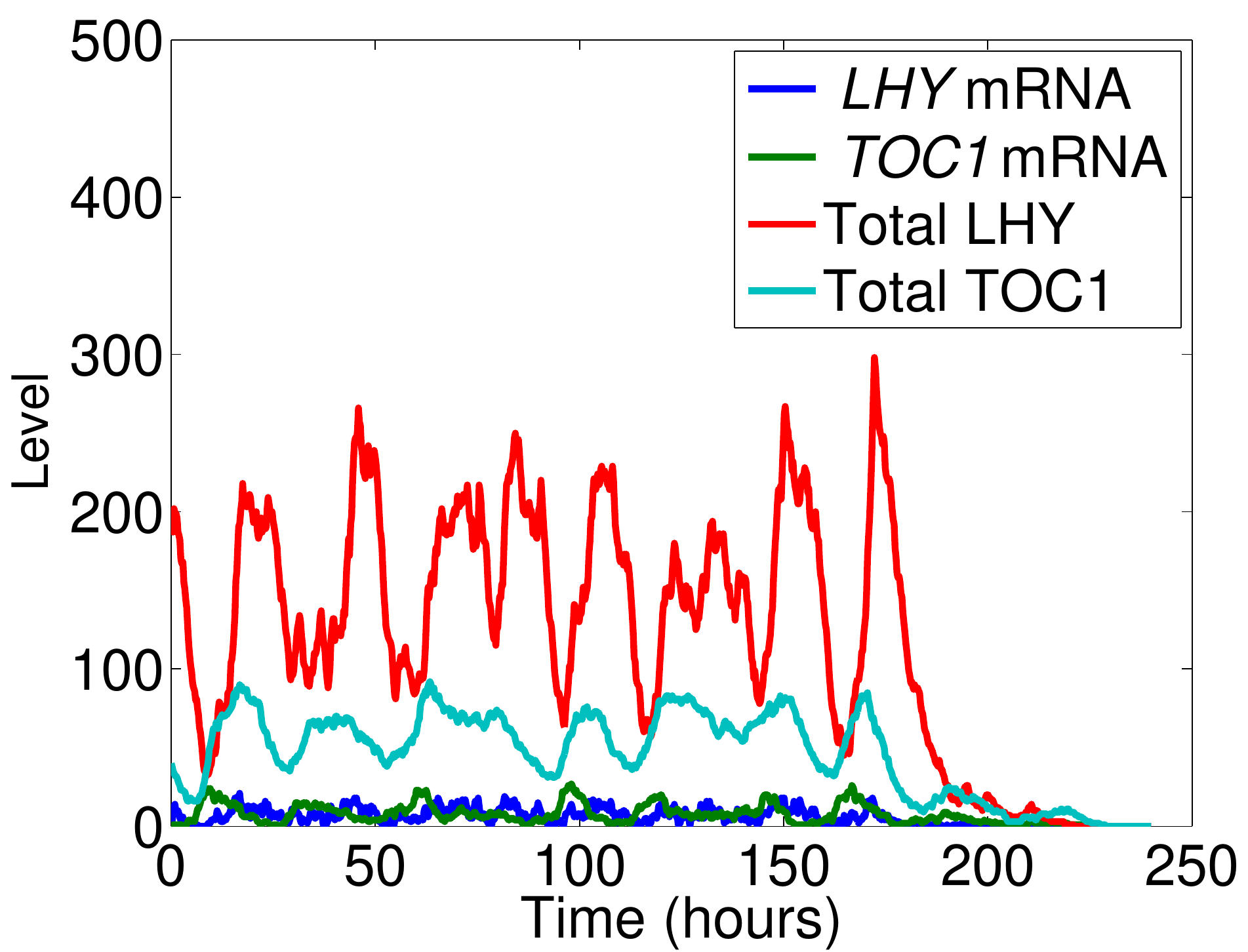}\label{fig:DD_LL_LLDD:LLDD_SSA1}}\vspace*{-.6ex}
  \caption{Comparison of the deterministic and stochastic models in different light conditions.}\label{fig:DD_LL_LLDD}
\end{figure}

\begin{figure}[hbt]
\centering
\subfigure[LD 6:18 -- ODE]{\includegraphics[width=0.32\textwidth]{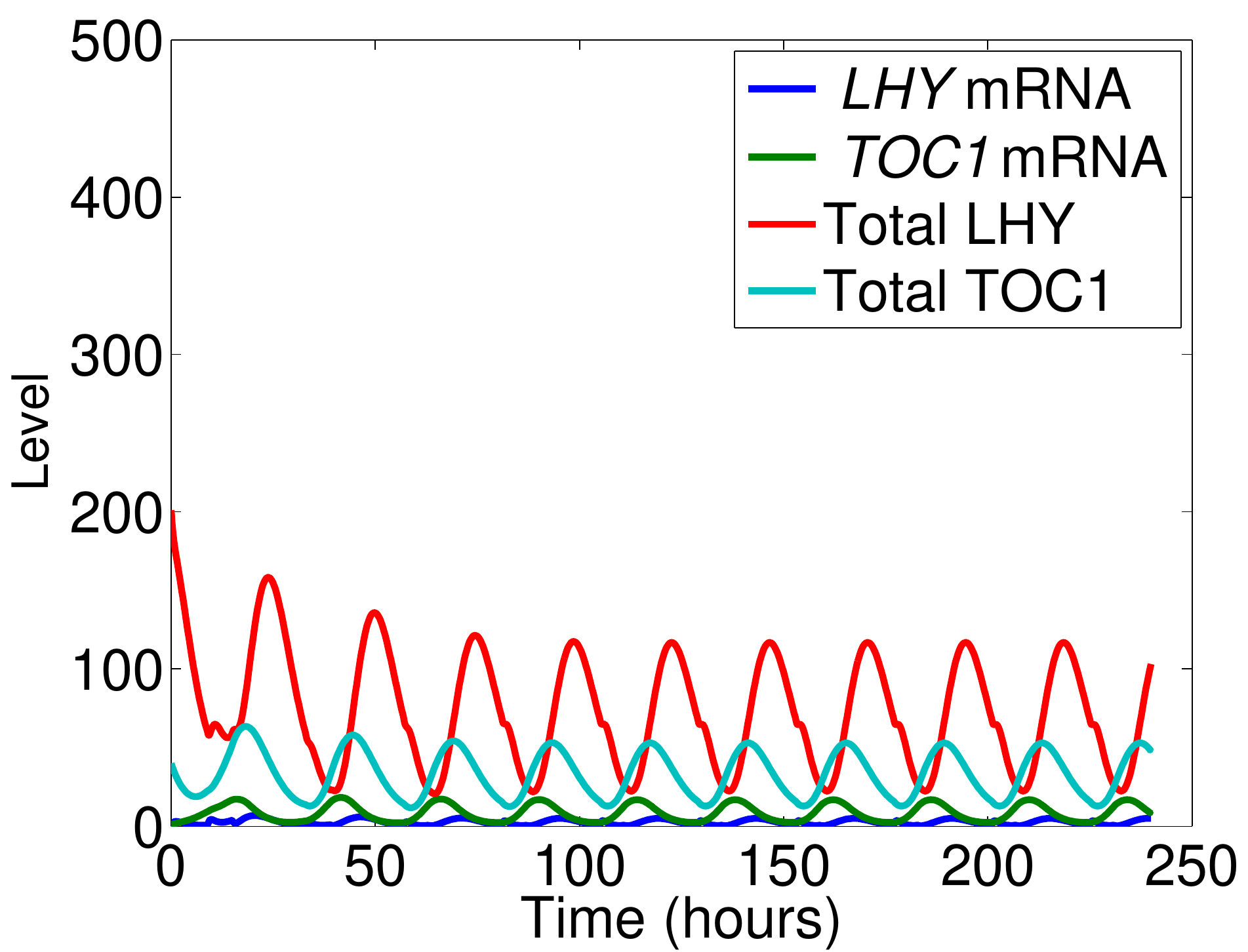}\label{fig:LD_all:6_18_ODE}}
\subfigure[LD 6:18 -- average 10000 SSA runs]{\includegraphics[width=0.32\textwidth]{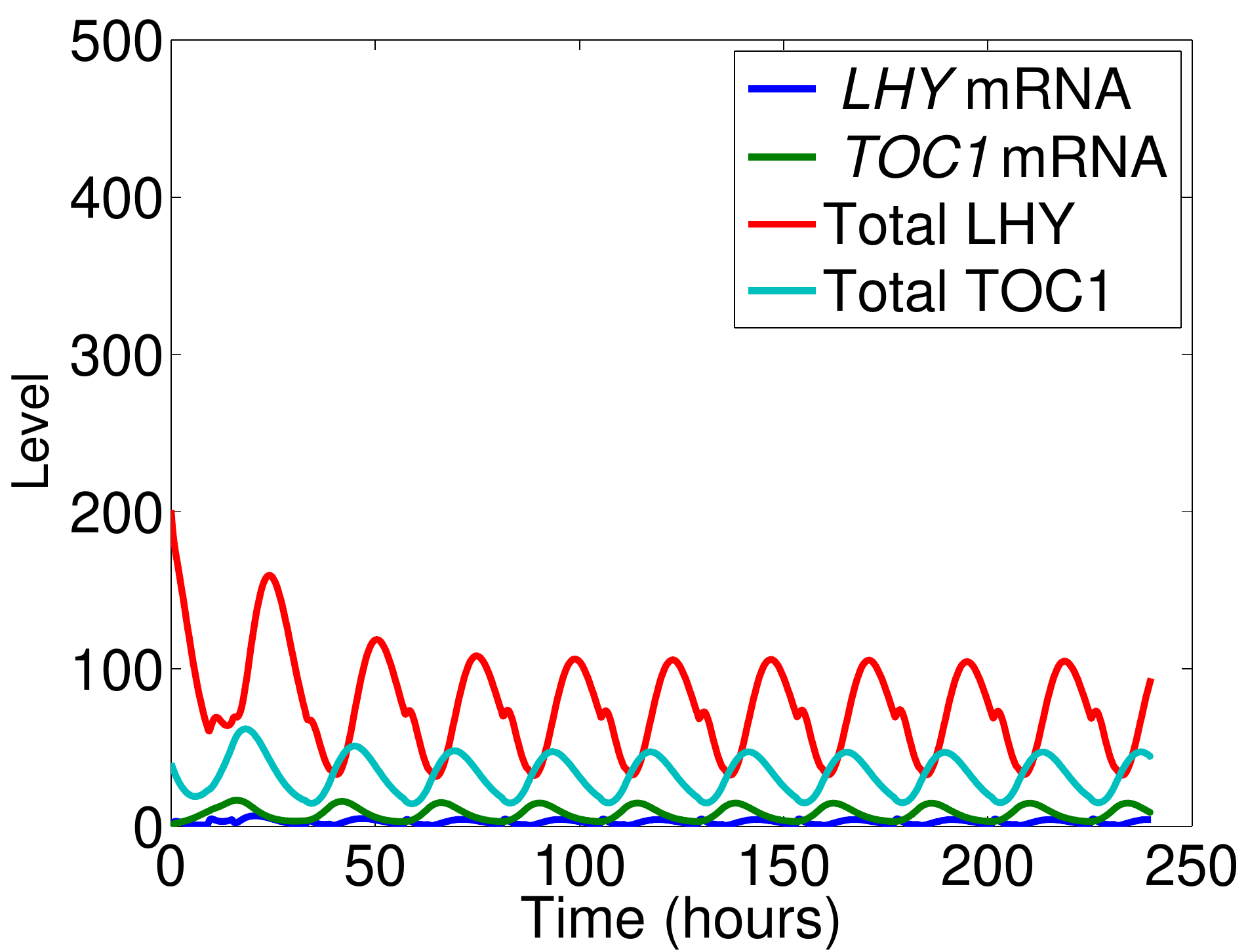}\label{fig:LD_all:6_18_SSA10000}}
\subfigure[LD 6:18 -- single SSA run]{\includegraphics[width=0.32\textwidth]{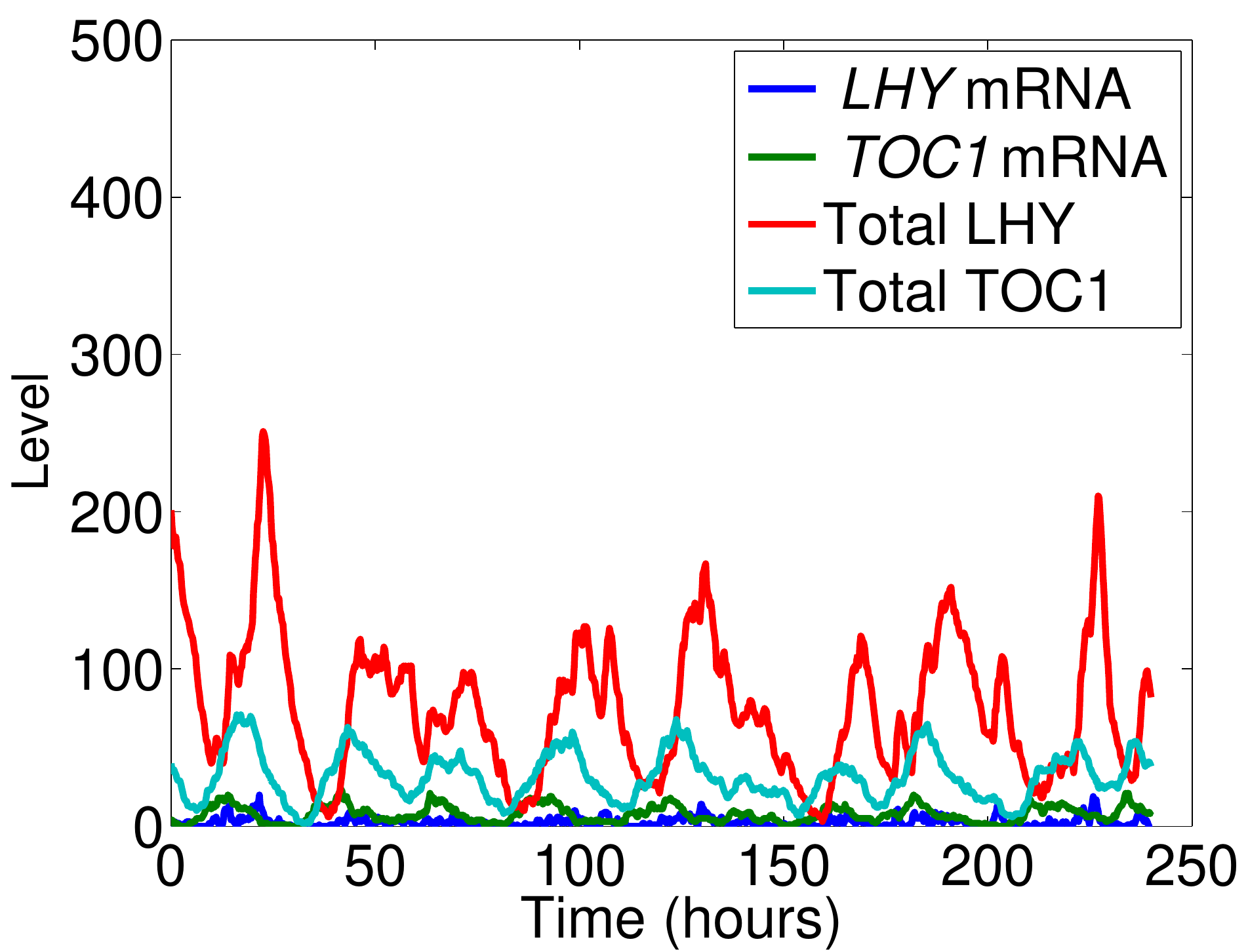}\label{fig:LD_all:6_18_SSA1}}\\\vspace{-1ex}
\subfigure[LD 12:12 -- ODE]{\includegraphics[width=0.32\textwidth]{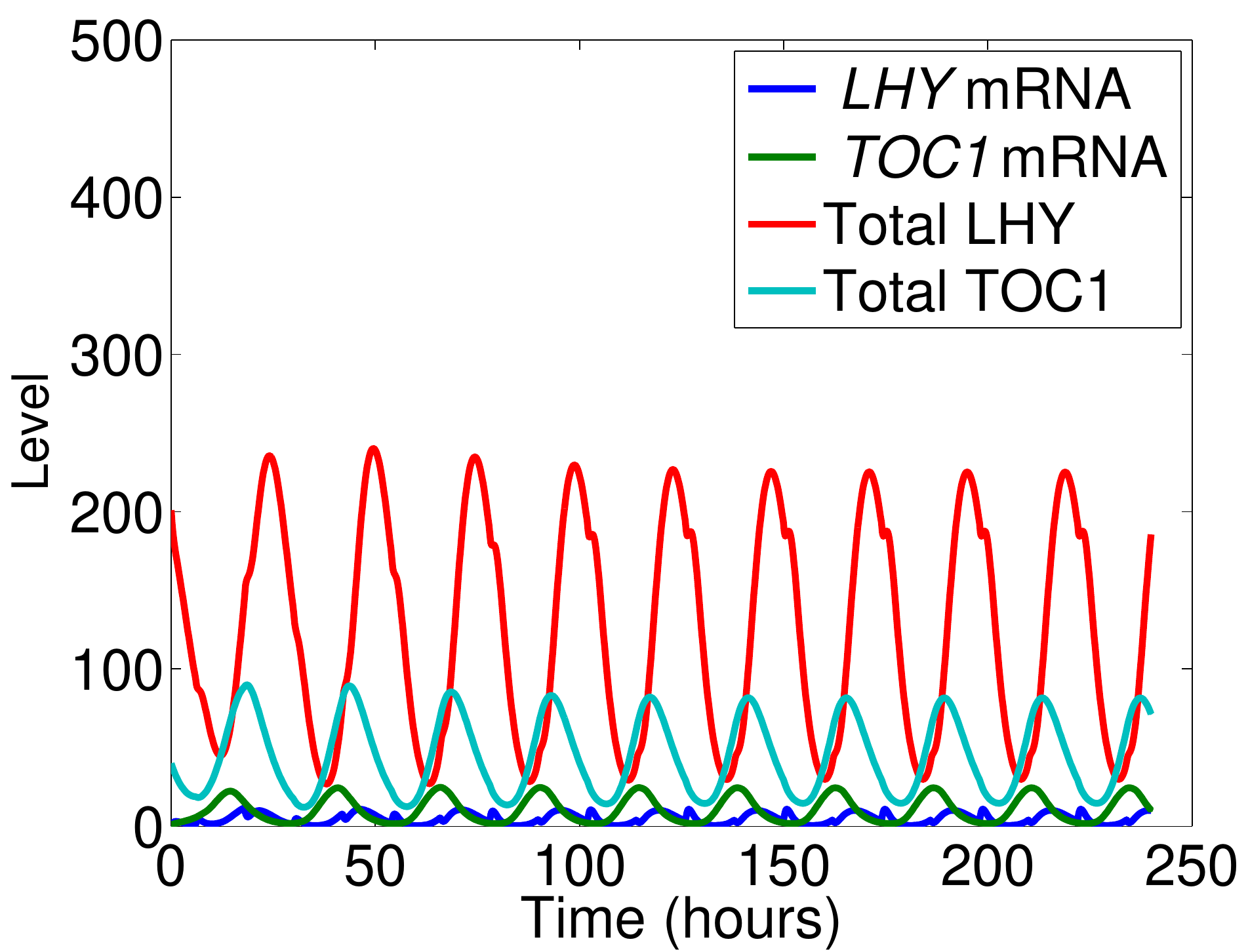}\label{fig:LD_all:12_12_ODE}}
\subfigure[LD 12:12 -- average 10000 SSA runs]{\includegraphics[width=0.32\textwidth]{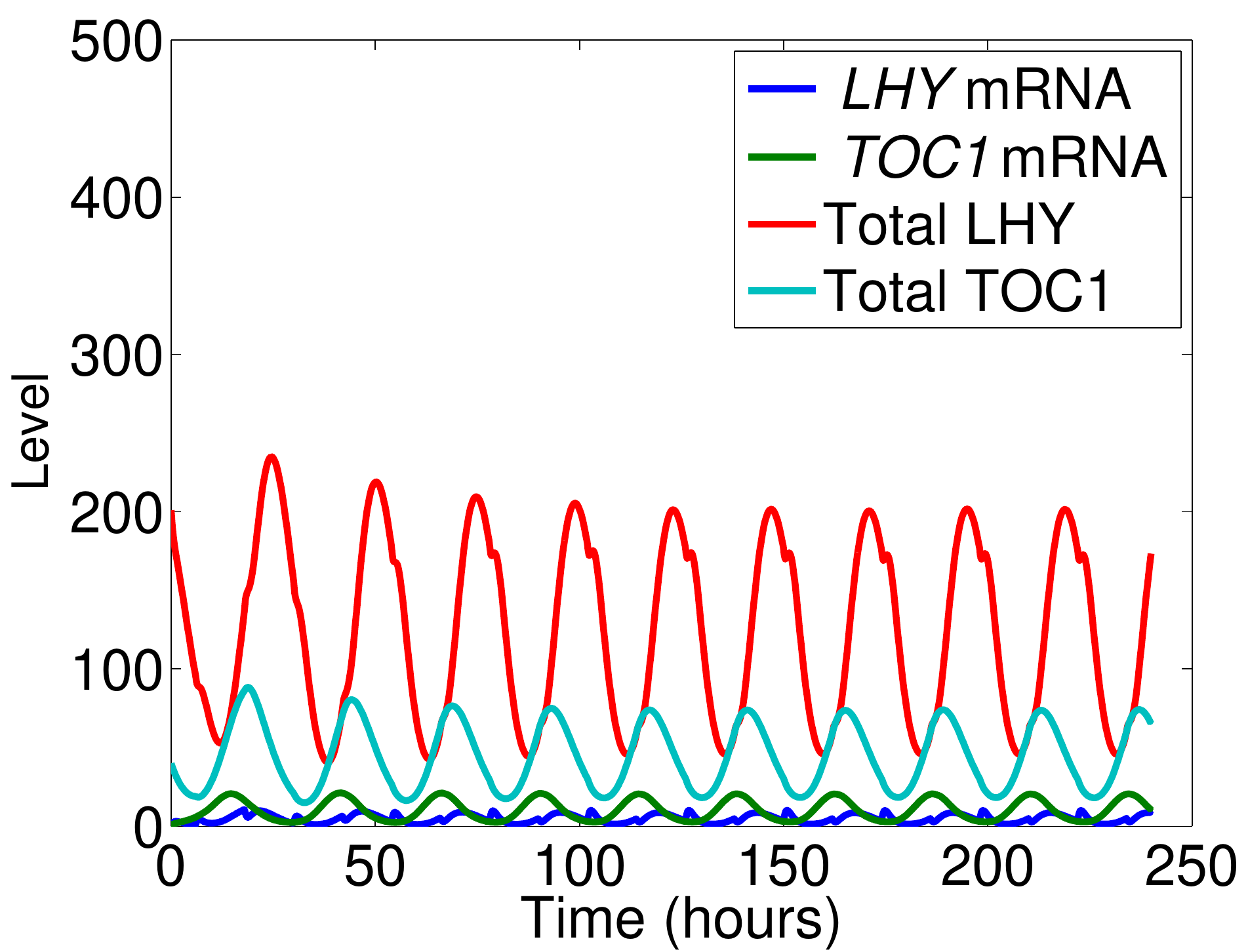}\label{fig:LD_all:12_12_SSA10000}}
\subfigure[LD 12:12 -- single SSA run]{\includegraphics[width=0.32\textwidth]{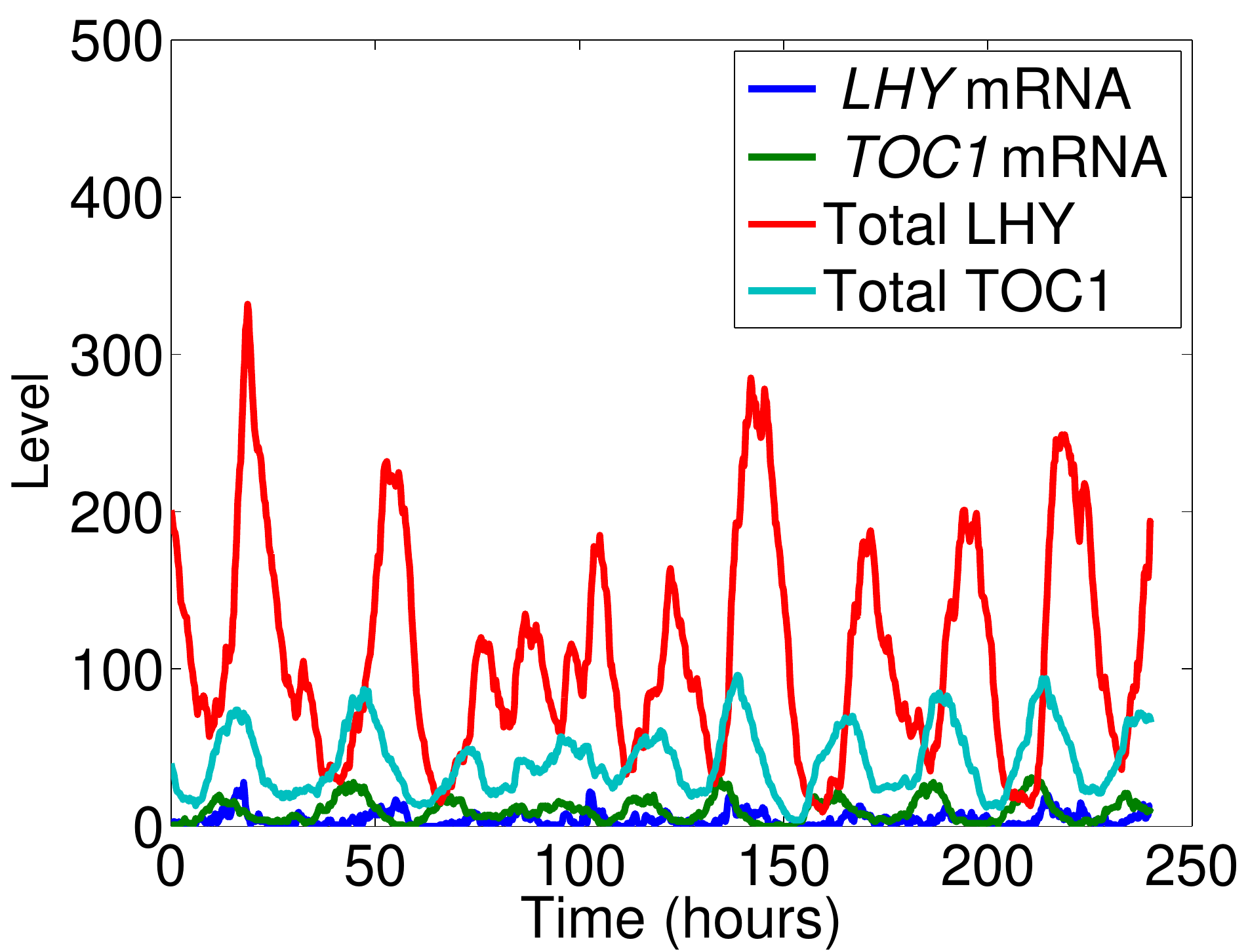}\label{fig:LD_all:12_12_SSA1}}\\\vspace{-1ex}
\subfigure[LD 18:6 -- ODE]{\includegraphics[width=0.32\textwidth]{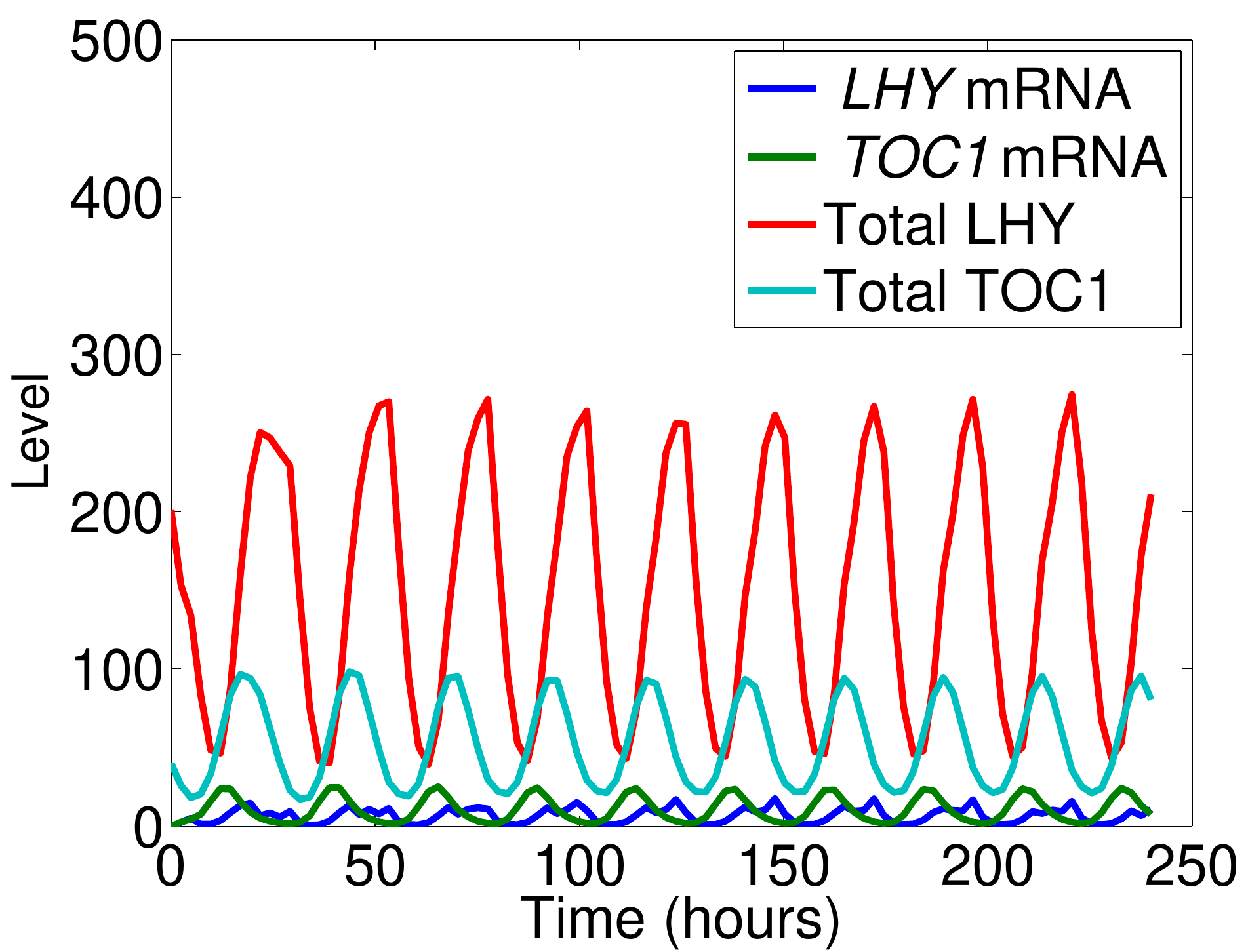}\label{fig:LD_all:18_6_ODE}}
\subfigure[LD 18:6 -- average 10000 SSA runs]{\includegraphics[width=0.32\textwidth]{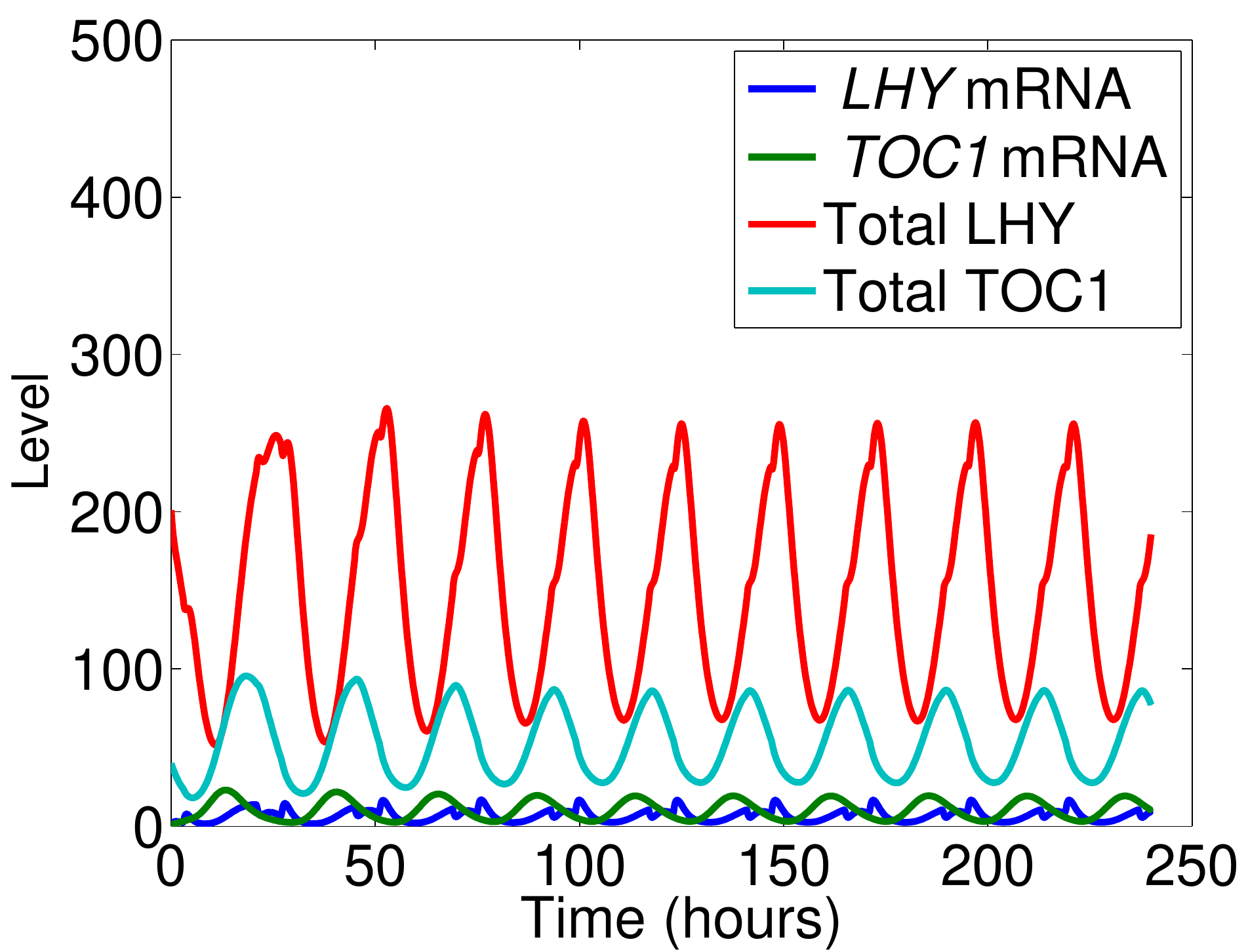}\label{fig:LD_all:18_6_SSA10000}}
\subfigure[LD 18:6 -- single SSA run]{\includegraphics[width=0.32\textwidth]{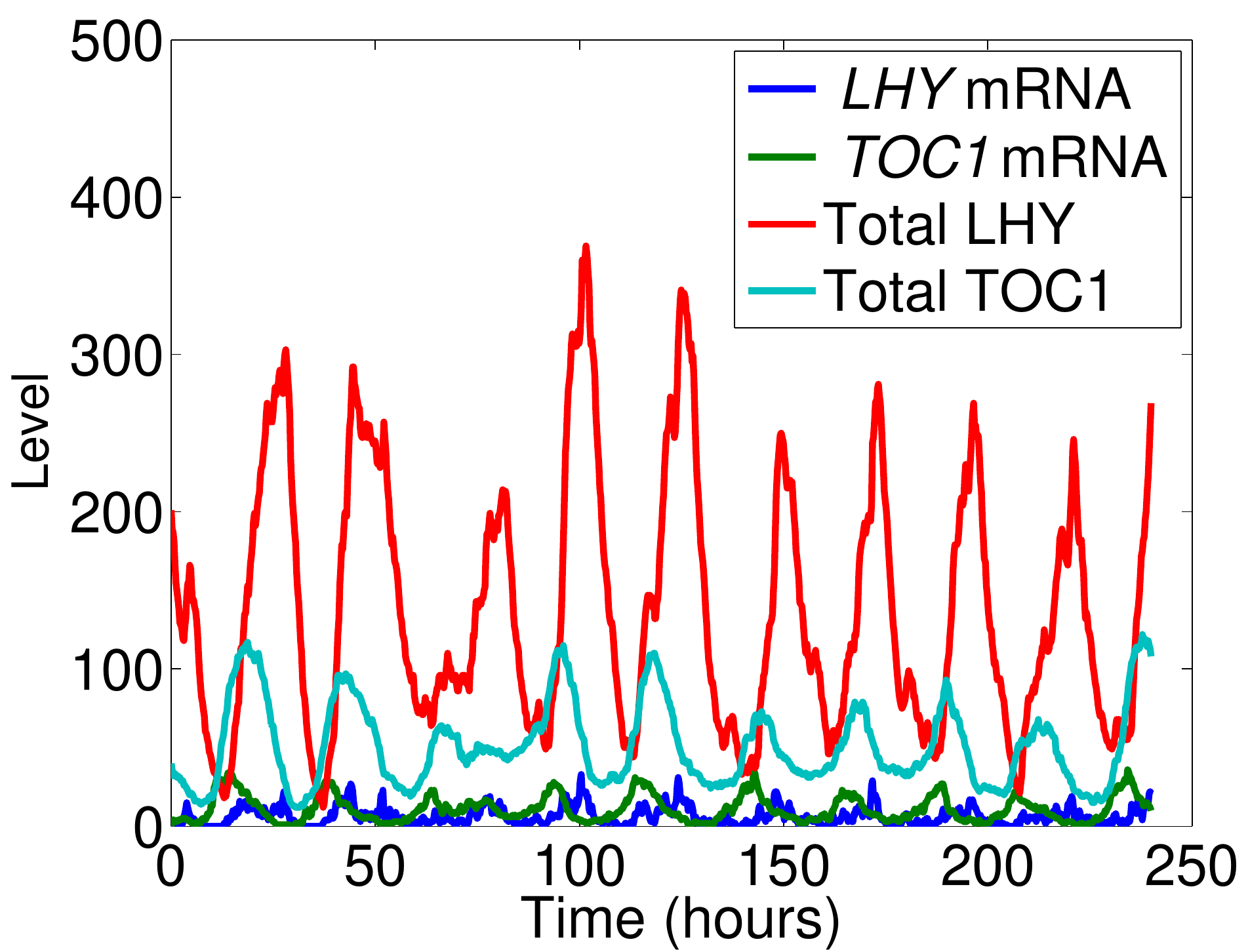}\label{fig:LD_all:18_6_SSA1}}\vspace{-.5ex}
  \caption{Comparison of the deterministic and stochastic models for different photoperiods.}\label{fig:LD_all}
\end{figure}

In the rest of this section we report results obtained
by solving the clock model using the Dormand--Prince ODE solver and
the Gibson--Bruck SSA, both available in
the Bio-PEPA Eclipse Plug-in~\cite{biopepa_site}.
%
%From here on we consider molecule count based ODE models that apply the same scaling as in SSAs in order to make results comparable.
%After such rescaling the  Bio-PEPA model can be analysed by ODEs and SSAs.
%
We consider three different light conditions: constant dark (DD),
constant light (LL), and alternating light/dark cycles (LD).
We also consider an experiment in which the system is transferred
from constant light into constant dark (LL-DD).
For each of these,
we compare results obtained by numerical integration of the
deterministic model with those obtained by stochastic
simulation.
The initial conditions are those of the original model at dawn following entrainment to 24
hour light/dark cycles (LD 12:12).
Figures~\ref{fig:DD_LL_LLDD} and \ref{fig:LD_all} report the
computed time-series behaviours for all settings.

The species of interest are \textit{TOC1\_mRNA},
\textit{LHY\_mRNA} and the corresponding experimentally observable total protein amounts
(\textit{Total\_TOC1} and \textit{Total\_LHY} as defined in
Section~\ref{sec:model:biopepa}).
%
% LL: I wonder whether the following sentence is still needed in the light of all explanations above.
%In order to compare the ODE and SSA results, we
%consider the ODE simulations in terms of molecule numbers
%(i.e. by rescaling concentrations by $\Omega$).

\vspace*{-1.5ex}
\paragraph{DD system.}

The rapid damping behaviour of the system in constant dark (DD) is shown in
Figure~\ref{fig:DD_LL_LLDD:DD_ODE}--\ref{fig:DD_LL_LLDD:DD_SSA1}. This
damping is seen in all analysis methods: ODEs, individual SSA
runs, and the mean stochastic behaviour calculated over 10000 independent
SSA runs. Despite the fact that the self-sustained oscillations observed by
averaging over the SSA runs stop very quickly (within 1-2 days), when
looking at individual SSA runs we note that occasionally a non-zero number
of molecules can be briefly observed, even after several circadian cycles
(Figure~\ref{fig:DD_LL_LLDD:DD_SSA1}).

%All species damp rapidly for both the solution of the deterministic system
%(Figure~\ref{fig:DDvsLLvsLD_all}(a)) and the solution obtained from averaging
%over 10000 independent runs of the stochastic model (Figures~\ref{fig:DDvsLL}(c)). In both cases, self-sustained
%oscillations are absent within 3-4 days. When
%looking at individual SSA runs we note that occasionally a
%non-zero number of molecules can be briefly observed even at later
%times (Figure~\ref{fig:DDvsLL}(e)).

\vspace*{-1.5ex}
\paragraph{LL system.}

In constant light (LL), the deterministic system
also exhibits damped oscillations, with all species tending to non-zero
constant values after about 7-8 days (Figure~\ref{fig:DD_LL_LLDD:LL_ODE}). Similar behaviour
is observed by averaging over 10000 SSA runs
(Figure~\ref{fig:DD_LL_LLDD:LL_SSA10000}), though the oscillations damp more
rapidly and the steady-state value is slightly different (e.g.~the
LHY copy number is about 130 in ODEs compared to about 160 in the SSA).
However, a 10-fold increase of the scaling factor to $\Omega=500$ yields a perfect
quantitative agreement between the SSA and ODEs (see Figure~\ref{fig:omega}
in Appendix~\ref{sec:appendix:results}). Although the precise source of the discrepancy
is not known at present, we hypothesise that it is caused by a breakdown of the continuous
approximation underpinning ODEs, due to the low copy numbers
obtained at small $\Omega$ values.
%ML: slightly rephrased the previous sentence

The other, most notable, difference between the deterministic and stochastic
models is the behaviour of individual SSA runs (Figure~\ref{fig:DD_LL_LLDD:LL_SSA1}), for which
persistent irregular oscillations (in both phase and amplitude) are observed.
Because of phase diffusion effects, however, these oscillations cannot be
detected in the mean behaviour: hence, in this case, neither the simple
average over multiple SSA runs nor the ODE solution gives us a correct
indication of the real behaviour of the system,
%validating the use of multiple analysis techniques in this case.
emphasising the importance of observing single realisations.
%ML: ok?

In view of these findings, we hypothesise that single cell
% (or even small population sizes)
experimental data may exhibit sustained oscillations, whereas the behaviour
observed in a large population of cells would be closer to the rapid damping reproduced by both the
ODE and SSA average. This is due to the fact that free-running oscillations are unlikely to be synchronised over different cells.
Furthermore, visual inspection of the solutions obtained by averaging over different numbers of SSA runs suggests
that it should be possible to discern the stochastic effects with a population of around 100 cells. The
experimental data currently available for this system are at the
level of a population comprising at least 10000 cells. However, we anticipate that
the development of new experimental techniques for measuring gene expression
in single cells or small populations will enable this hypothesis to tested
experimentally in the not-too-distant future.

%If populations of decreasing size are observed, then our simulations
% predict that damping effects are expected to get weaker for smaller
% populations.
%ML: Last sentence seemed a repetition

\vspace*{-1.5ex}
\paragraph{LL-DD system.}

As an additional experiment, we consider a system which is kept in constant
light (LL) for 160 hours, and then transferred into constant dark (DD). The
time-series results are reported in Figures~\ref{fig:DD_LL_LLDD:LLDD_ODE}--\ref{fig:DD_LL_LLDD:LLDD_SSA1}.
It can be seen that single realisations of the SSA --- approximating the
behaviour of individual cells --- exhibit immediate cessation of
self-sustained oscillations following the LL-DD transfer. This behaviour can
be understood by considering the fixed points of the deterministic model.

\begin{figure}[hbt]
\centering
\includegraphics[width=0.49\textwidth]{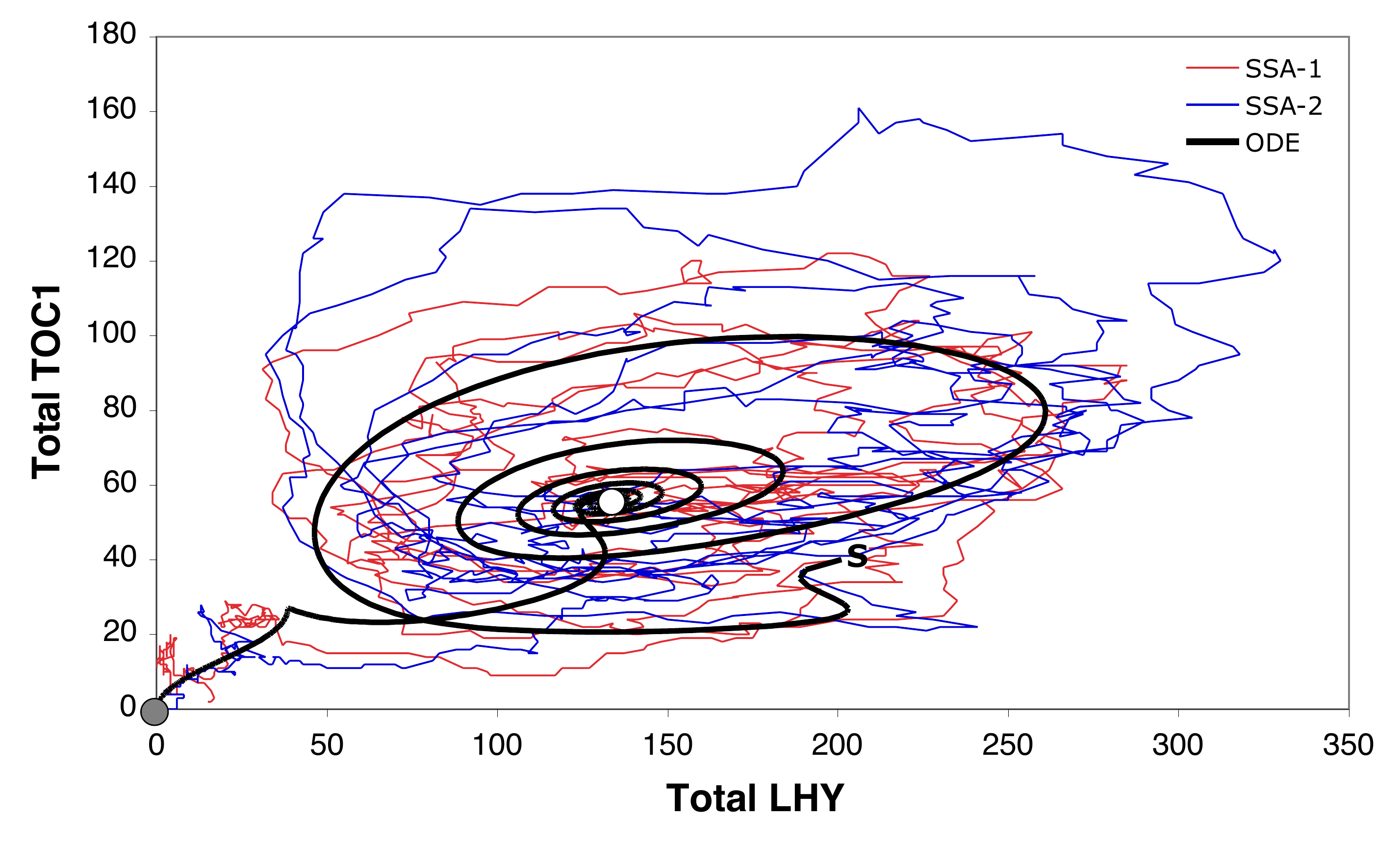}\vspace*{-.5em}
  \caption{Phase-space for the LL-DD transfer experiment. White and grey
  dots indicate the fixed points (steady-states) of the deterministic system
  in LL and DD conditions, respectively. The eigenvalues of the model obtained by
  linearising the deterministic equations around a fixed point determine the behaviour of the system in
  its neighbourhood. These are listed for both fixed points in Table~\ref{tab:evalues:LLDD} of Appendix~\ref{sec:appendix:results}.}\label{fig:LLDD_phasespace}
\end{figure}

The LL fixed point is located far from the origin in phase space and is of the stable focus type. Trajectories
of the ODE --- approximating the behaviour of a large population of cells --- spiral around the
fixed point as they converge to it, producing slowly damping oscillations. In the corresponding
stochastic model, fluctuations kick individual realisations of the system between these spiralling
trajectories, thereby preventing the system from remaining close to the
fixed point for long periods (see Figure~\ref{fig:LLDD_phasespace}). This leads to the irregular self-sustained
oscillations observed. By contrast, the DD fixed point of the ODE system
is located at the origin. As species concentrations must be positive, the fixed point cannot be a stable
focus, and is instead a stable node. Trajectories of the ODE converge directly onto it, generating
oscillations that quickly damp to zero. Individual
realisations of the stochastic model are thus repeatedly perturbed between rapidly convergent
trajectories. Consequently, they quickly approach
the DD steady-state following the LL-DD transition, remaining in its vicinity
thereafter (see Figure~\ref{fig:LLDD_phasespace}).

The model thus predicts that the DD behaviour of the LL-DD system at
the single-cell level mirrors that at the population level. If
further experiments were to reveal that self-sustained oscillations
are observed during the DD phase, this would indicate that the model
requires modification to convert the DD fixed point into a stable
focus bounded away from the origin.

\vspace*{-1.5ex}
\paragraph{LD system.}

The light conditions considered so far are experimental settings useful
for observing the system's endogenous dynamics. It
is also informative, however, to observe the behaviour of the clock under
natural conditions (alternating 24-hour cycles of light and dark). We present
results obtained for three different photoperiods: 6 hours
light/18 hours dark (LD 6:18), 12 hours light/12 hours dark (LD
12:12), and 18 hours light/6 hours dark (LD 18:6).

As described in Section~\ref{sec:model}, exposure to periodic
external stimuli such as light/dark cycles has the effect of
resetting the free-running oscillations observed in constant light,
so as to establish stable phase relationships with the forcing
stimulus.
Compared with the free-running LL system, the entrainment to LD
cycles regularises the dynamics of the system, markedly reducing the
variability of oscillations, particularly in terms of phase. As a
consequence, persistent regular oscillations with a stable phase
relationship to the light/dark cycle are observed in both ODEs
(Figures~\ref{fig:LD_all:6_18_ODE}, \ref{fig:LD_all:12_12_ODE},
\ref{fig:LD_all:18_6_ODE}) and when averaging over multiple SSA runs
(Figures~\ref{fig:LD_all:6_18_SSA10000},
\ref{fig:LD_all:12_12_SSA10000}, \ref{fig:LD_all:18_6_SSA10000}).
This phase regularisation can also be seen in individual SSA runs of
the entrained system (compare Figures~\ref{fig:LD_all:6_18_SSA1},
\ref{fig:LD_all:12_12_SSA1}, \ref{fig:LD_all:18_6_SSA1}) with the
simulations of the free-running clock in
Figure~\ref{fig:DD_LL_LLDD:LL_SSA1}). These observations are
consistent with previous stochastic analyses of clock
models~\cite{Gonze02,akmanEtAl09_neurospora_clock_biopepa}. We also
note that, as for the LL system, the deterministic and mean
stochastic behaviour, whilst very similar, are not in perfect
agreement.

\vspace*{-1ex}

\subsection{Model-checking: time-dependent probability distributions}
\label{sec:methods_results:mc}

Model-checking~\cite{clarke-grumberg-peled99} is a formal
verification method that allows modellers to state properties of a
given model and to automatically check whether they are met. Probabilistic
model-checking (see, for instance,
\cite{baierEtAl03,hintonEtAl06}) adds probabilistic measures in the
evaluation of queries. Recently, model-checking has grown in
popularity within the field of systems biology due to its ability to
directly answer complex questions on a model's behaviour. Traditional model-checking
verification differs from simulation-based analysis in that the verification of a
property is obtained from a computation over the entire
state-space of the continuous-time Markov chain (CTMC) underlying
the model. The major drawback of this approach is
the state-space explosion problem: the
model's dimension is often too large for computational viability.

Statistical model-checking (see, for instance,
\cite{younes-simmons02}) is an alternative query-based approach: it
estimates the probability distributions and computes approximate
results of queries (together with an estimate of the
%potential
error) by generating
%a given number of
random realisations of the CTMC and averaging the results obtained
by evaluating the queries on each of them. The advantage of
statistical model-checking over exact verification approaches is
that it does not need to build the explicit state-space of the
model, which is often intractable, and it does not rely on the
transient solution of the CTMC.
%Moreover, in probabilistic
%model-checking, the computation of results often involves the
%transient solution of the CTMC which is often infeasible.
%Statistical model-checking does not need this, which makes the
%analysis much faster or often even simply feasible.
%
%
%The obvious drawback of statistical model-checking compared to exact
%verification is that any probabilities that are computed by this approach
%cannot be smaller than $1/n$, where $n$ is the number of observed
%simulations.
%ML: I think this is not entirely correct, reverted to previous sentence
%
%MLG: added the following sentence
In essence, statistical model-checking is a verification technique
which allows modellers to perform additional analyses of a stochastic
system by automatically evaluating queries over multiple simulation
traces.
The obvious drawback of statistical model-checking
%compared to exact verification
is that it only considers a finite number of behaviours
of the system (i.e.~paths in the CTMC) and, hence, the accuracy of the results is strongly
related to that number.
However, exact verification of
biological systems is generally infeasible, and statistical
model-checking often represents a good practical solution.
Another issue of probabilistic model-checking is that the transient
solution of the CTMC can incur the same averaging effect discussed
previously relating to ODEs and mean SSA behaviour: computing the
expected value of the model variables might not be sufficient
because this would be exactly the same as the deterministic
behaviour. Results of reward-based properties, for instance, are
computed in terms of expected values and this, especially in the
case of oscillations which are out of phase, does not give
satisfactory results.

PRISM~\cite{hintonEtAl06} is a probabilistic model-checker, which
can be used to verify properties of a CTMC model. It also includes a
discrete-event simulator
%which can be used to perform
for statistical
model-checking. PRISM has been used to analyse systems from a wide
range of application domains, and recently also biochemical systems~\cite{heathEtAl08}.
Models are described using the state-based PRISM language, and it is
possible to specify quantitative properties of the system using a
property specification language which includes the temporal logic
CSL (Continuous Stochastic
Logic)~\cite{azizEtAl96,baierEtAl03}.
%the temporal logics
%CSL~\cite{azizEtAl96,baier-katoen-hermanns99}, LTL, PCTL, and
%PCTL$^*$~\cite{hansson-jonsson89}, and extensions for quantitative
%specifications and costs/rewards.

%\subsubsection{Application to the \textit{Ostreococcus} clock}

%\paragraph{}

Using the Bio-PEPA Workbench~\cite{biopepa_site}, we generated a
PRISM model of the clock (together with a set of reward structures and some
standard CSL properties which are automatically generated).
In the PRISM model, one module is defined for each species, and
module local variables are used to record the current quantity of
each species. The transitions correspond to the activities of the
Bio-PEPA model and the updates take the stoichiometry into account.
Transition rates are specified in an auxiliary module which defines
the functional rates corresponding to all the reactions. In order to
have a finite CTMC, lower and upper bounds are defined for each
variable.
%(i.e.~for the amount of each species)
In the following, we
focus on statistical model-checking: in this case, the choice of the
values for the bounds has no effect on the performance of the
analysis (since the CTMC is not built) and so we can use arbitrarily
high values that are guaranteed not to be reached.

\paragraph{Modelling the light in PRISM.}

In order to model the entrained clock (LD system), time-dependent
events must be represented. However, because of the intrinsic nature
of model-checking algorithms, which involve the numerical solution
of the CTMC underlying stochastic models, deterministic events and
time-dependent functions cannot be explicitly specified in PRISM.

We investigated several approaches to address this problem. A first
possibility is to split the model-checking algorithm into different
(two or more, depending on the time window we are interested in)
analysis steps over different time intervals, each with constant
light conditions: two different CTMCs would be considered (one for
the day-time system and one for the night-time one) with the
algorithm switching back and forth between them. The main issue with
this approach is how to merge the results obtained over the
different time periods. For some particular queries, such as those
relating to reachability, this can be done by splitting them into a number
of subqueries such that the result (i.e.~probability distributions)
of one query can be used as the initial state for the next one. For
an arbitrary CSL query, however, this cannot be done, and this
strongly limits the kind of queries that could be verified.

The alternative approach we consider here is to represent the light
by approximating time using a monotonically increasing stochastic
variable. The main issue of this
approach is that we introduce an additional stochastic effect which
is absent in the system we have described so far (i.e.~where light
is modelled as a deterministic on/off switch). In practice this does
not matter, provided that the stochastic variability introduced is
kept smaller than the variability of any experiments the model may
be compared to.
%Provided one ensures that the introduced stochastic
%variability is not excessive, a claim in favour of modelling light
%as a stochastic function is that this representation is closer to
%what happens in real life (and even in the laboratory there is
%always a small experimental variability which might cause the
%process not to be perfectly deterministic).
The introduction of an
additional variable to model time also causes an increase in the
state-space, but this is not an issue here since we focus
on statistical model-checking only. An extract from the PRISM model
showing how we model time and the day/night switch is provided in
Appendix~\ref{sec:appendix:prism}.

\paragraph{Time-dependent probability distributions of protein levels.}

Using model-checking we can compute the time-dependent probability
distributions for each of the model species. For instance, by
verifying the CSL property
$$\Prob{\Eventually{[T,T]}( \mathit{LHY\_c} + \mathit{LHY\_n} = i)}$$
\noindent for time instant $T \in [0,96]$ and protein level $i \in
[0,500]$, we can observe how the probability distribution for LHY
protein changes over time during the first 96 hours of simulation.

\begin{figure}[hbt]
\centering
\subfigure[DD]{\includegraphics[width=0.48\textwidth]{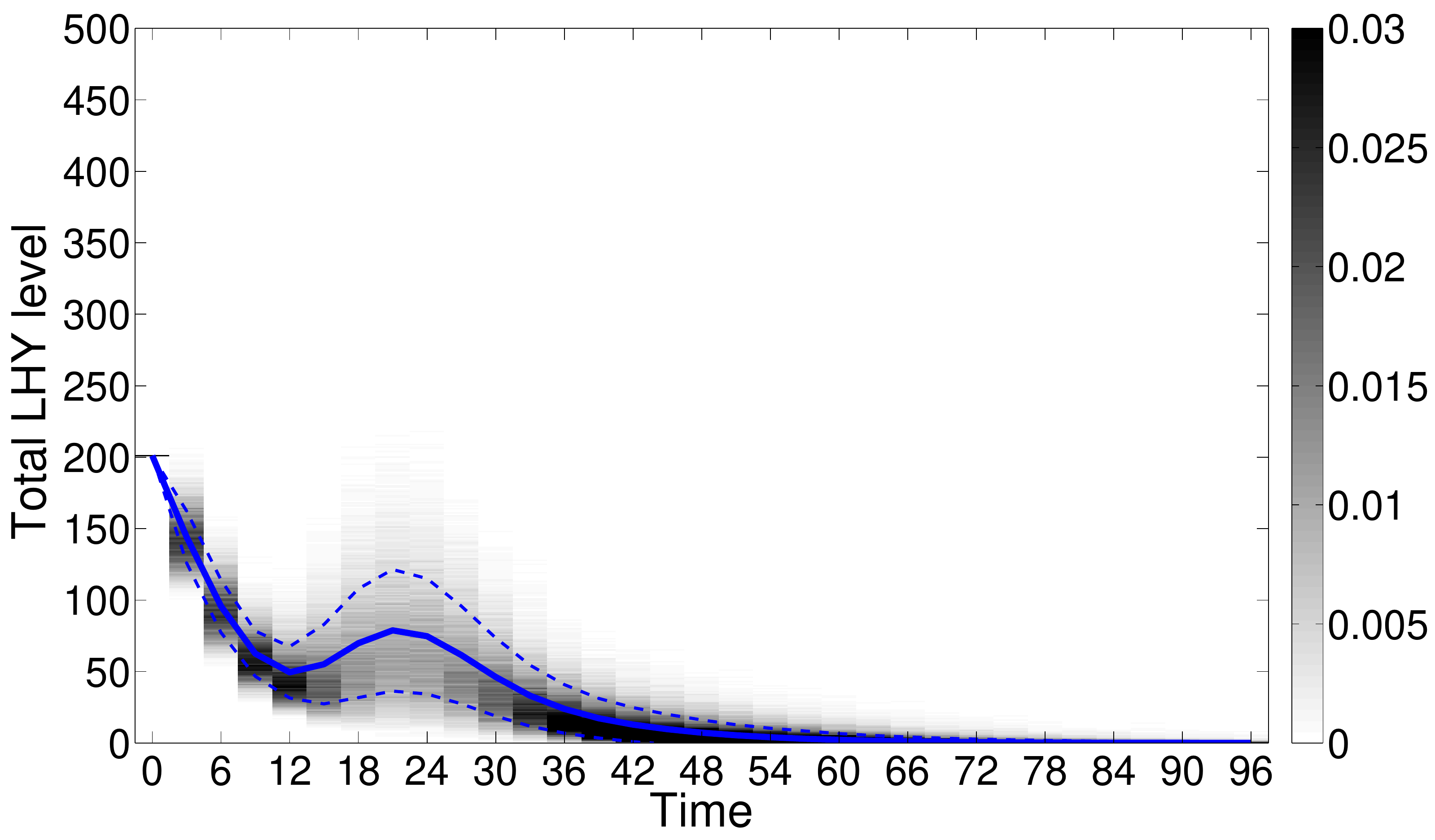}}\\\vspace{-1ex}
\subfigure[LL]{\includegraphics[width=0.48\textwidth]{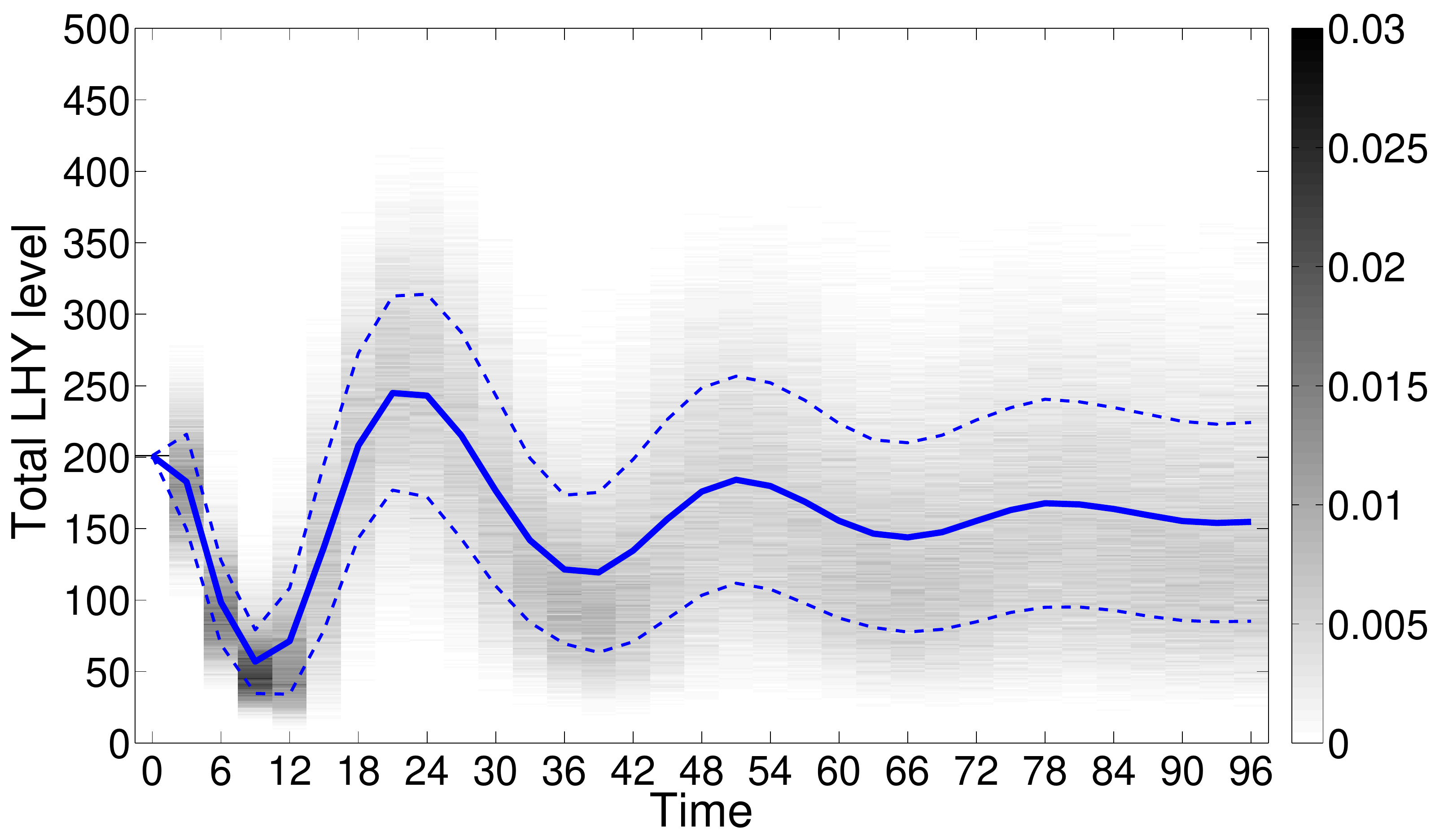}}
\subfigure[LD 12:12]{\includegraphics[width=0.48\textwidth]{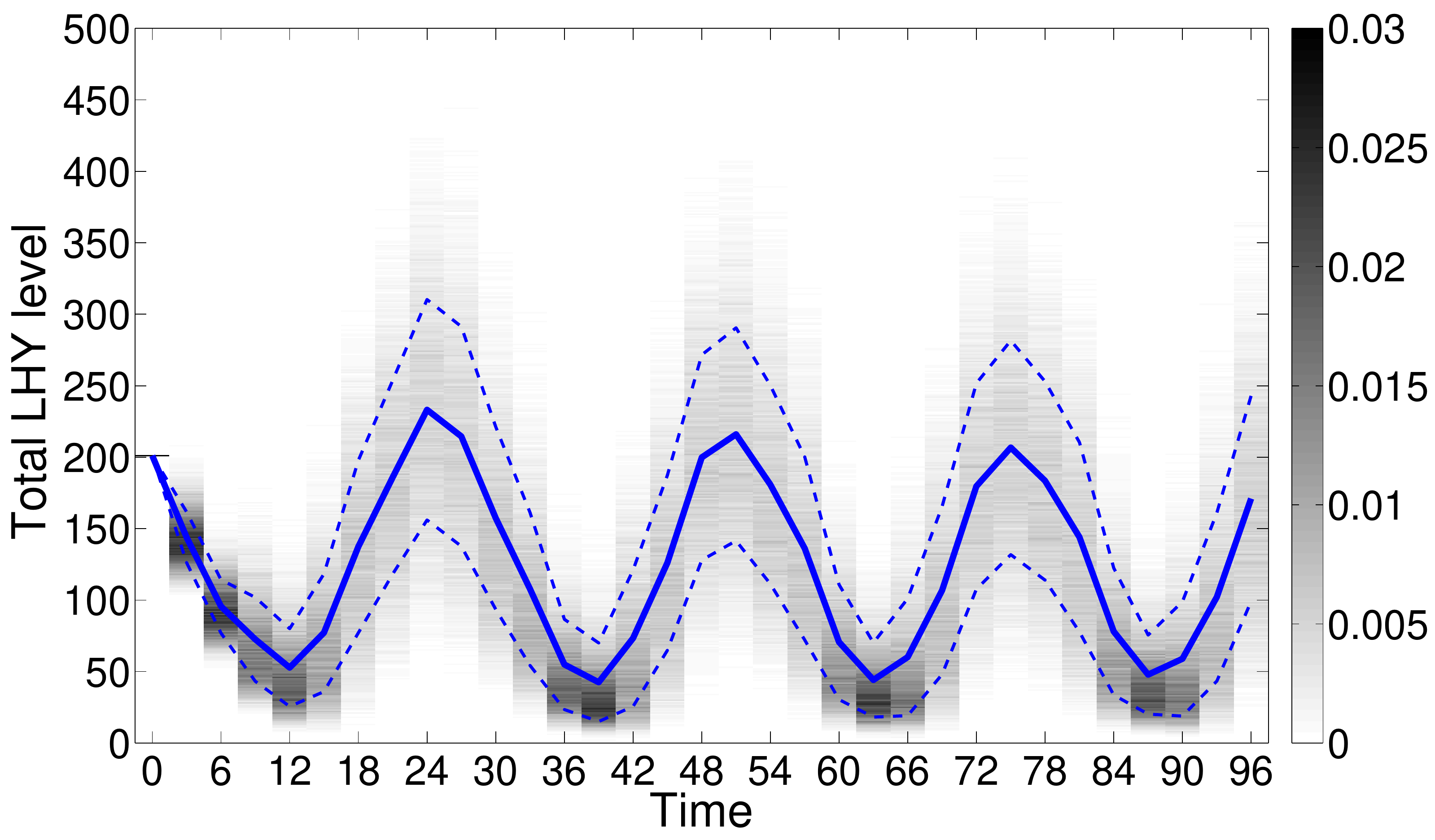}}\vspace{-.5ex}
  \caption{Probability distribution of total LHY level over the first 4 days (0--96 hours, 10000 runs). The heatmap represents the probability distribution: a darker colour corresponds to a higher probability. Blue lines show average and standard deviation ($\mu \pm \sigma$). Similar changes in distribution with time were
  obtained for TOC1 protein.}\label{fig:mc_lhy_distribution_4days}
\end{figure}

Each of the plots in Figure~\ref{fig:mc_lhy_distribution_4days}
refers to a different light condition (DD, LL, and LD 12:12),
showing how the probability of being at a particular level changes
over time in each case. The plots also report the mean value $\mu$
and the standard deviation $\sigma$ of LHY expression. In all cases,
the initial amount is $\mathit{LHY} = 200$, and then the probability mass
gradually moves away from this initial value. In DD, as expected
from the results of Section~\ref{sec:methods_results:ssa}, the bulk of the probability
mass rapidly moves close to zero. In LL we can clearly
observe the effect of phase diffusion, resulting in a probability
distribution spread almost equally across a wide range of values. By
contrast, in LD we are able to observe clear oscillations in the
probability distribution; we also note that the amplitude of peak
expression is more variable than that of trough expression,
with a much broader spread of the probability mass around the mean.

%\begin{figure}[hbtp]
%\centering
%\subfigure[DD]{\includegraphics[width=0.5\textwidth]{Figs/DD_P_Toc1_equal_i_at_time_T_0_3_96_10000run_heatplot_avg_std.pdf}}\\\vspace{-2ex}
%\subfigure[LL]{\includegraphics[width=0.5\textwidth]{Figs/LL_P_Toc1_equal_i_at_time_T_0_3_96_10000run_heatplot_avg_std.pdf}}\hspace{-3ex}
%\subfigure[LD 12:12]{\includegraphics[width=0.5\textwidth]{Figs/LD_P_Toc1_equal_i_at_time_T_0_3_96_10000run_heatplot_avg_std.pdf}}
%  \caption{Probability distribution of total TOC1 level over the first 4 days (0--96 hours) (10000 runs).}\label{fig:mc_toc1_distribution_4days}
%\end{figure}

\begin{figure}[hbt]
\centering
\subfigure[LHY -- probability distribution and $\mu \pm \sigma$]{\includegraphics[width=0.505\textwidth]{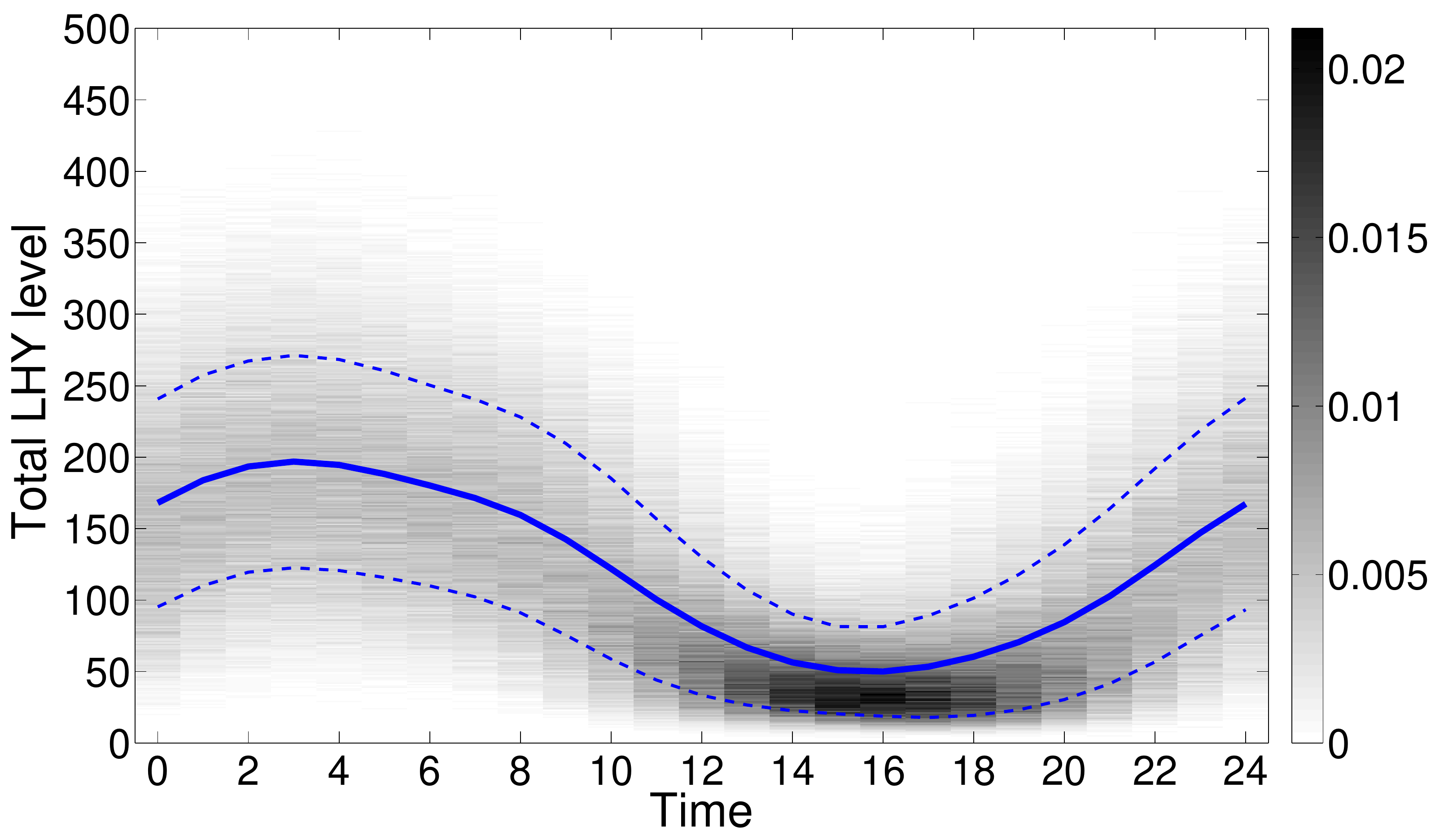}}
\subfigure[LHY -- coefficient of variation $c_v = \sigma/\mu$]{\includegraphics[width=0.48\textwidth]{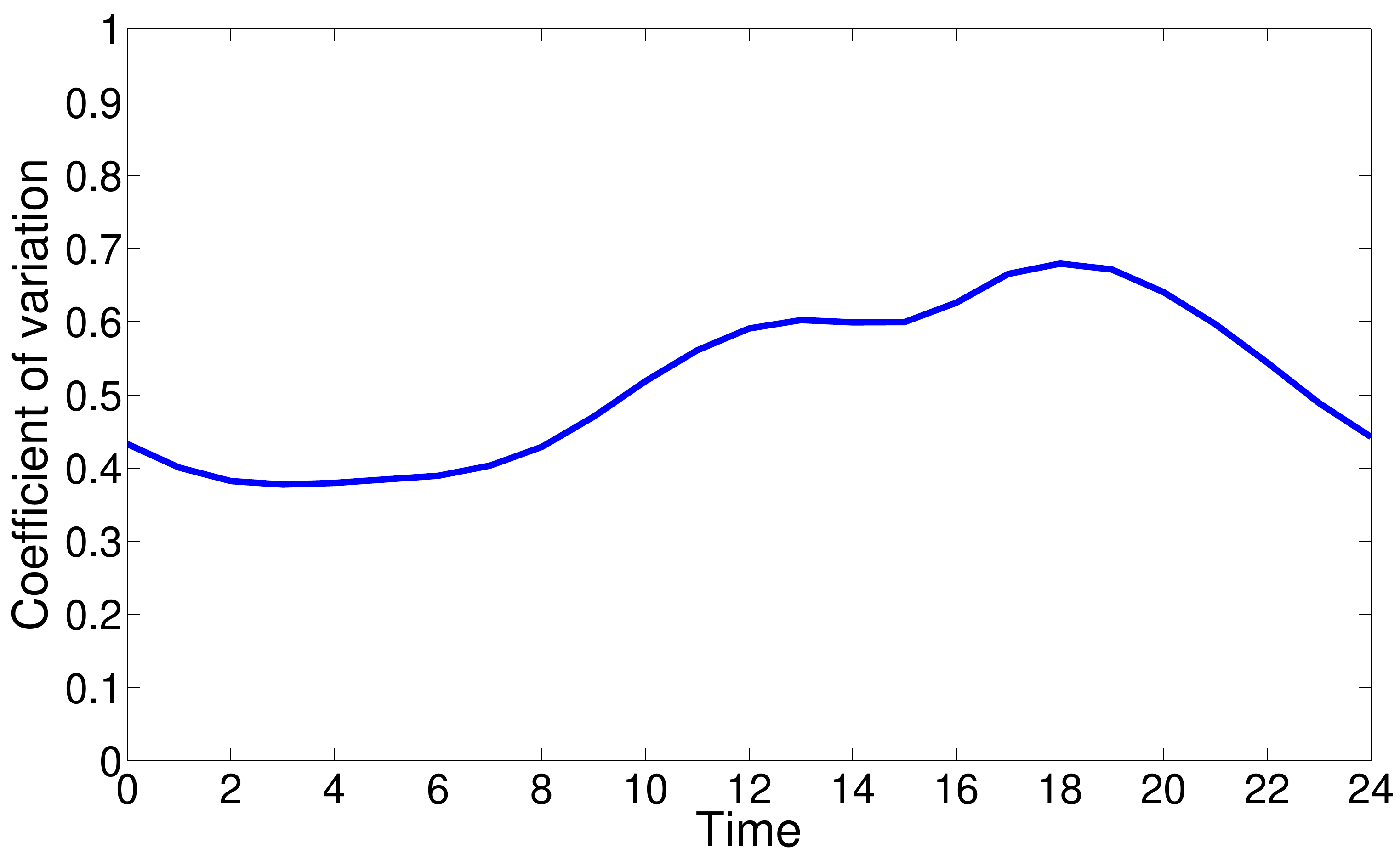}}\vspace{-.5ex}
%\subfigure[TOC1 -- distribution, $\mu \pm \sigma$]{\includegraphics[width=0.49\textwidth]{Figs/LD_P_Toc1_equal_i_at_time_T_10000run_heatplot_avg_std.pdf}}
%\subfigure[TOC1 -- $c_v = \sigma/\mu$]{\includegraphics[width=0.49\textwidth]{Figs/LD_P_Toc1_equal_i_at_time_T_10000run_coeff_var.pdf}}
  \caption{Probability distribution of LHY %(top) and TOC1 (bottom)
  level over one day (120--144 hours) in LD 12:12 (10000 runs). The distribution of TOC1 exhibits a qualitatively similar variation.}\label{fig:mc_distribution_1_day}
\end{figure}

In the following, we focus on the case of alternating light/dark
cycles (LD 12:12). In Figure~\ref{fig:mc_distribution_1_day}(a) we
report the probability distribution for LHY protein, together with
its mean $\mu$ and standard deviation $\sigma$ in a single 24-hour
cycle (from 120 to 144 hours).
Figure~\ref{fig:mc_distribution_1_day}(b) plots the oscillation in
the corresponding coefficient of variation $c_v=\frac{\sigma}{\mu}$.
This provides a normalised measure of the sensitivity of the LHY protein
oscillation to stochastic fluctuations as a function of
circadian time. Small values of $c_v$, therefore, correspond to robust
phase markers (a high signal-to-noise ratio) and large values
poor phase markers (a low-signal-to-noise ratio).
Commonly used phase measures in circadian research are the times of
peak and trough expression together with the time at which the
oscillation falls to its midpoint level. It can be seen in
Figure~\ref{fig:mc_distribution_1_day}(b) that the coefficient of
variation is minimal around the peak, suggesting that the latter is
the best of the standard phase markers for analysing experimental LHY data.

\begin{table}[b]
  \centering
\begin{small}
\begin{tabular}{|c|c|c|c|c|c|c|c|c|c|c|c|}
  \hline
  \textbf{\textit{e}}    & 0      & 2      & 4      & 6      & 8      & 10     & 12     & 14     & 16     & 18     & 20  \\
  \textbf{P} & 0.8768 & 0.9276 & 0.9545 & 0.9737 & 0.9833 & 0.9903 & 0.9931 & 0.9964 & 0.9987 & 0.9993 & 0.9998 \\
  \hline

\end{tabular}
\end{small}
\caption{Probability of total LHY to stay below the threshold $e$ in
DD (96--500 hours, 10000 runs).}\label{tab:DD:constant}
\end{table}

As discussed in Section~\ref{sec:methods_results:ssa}, for the DD
system, both the deterministic solution and SSA average quickly attain a constant value.
Species amounts can, however, be greater than zero for short time intervals in individual SSA runs (see
Figure~\ref{fig:DD_LL_LLDD:DD_SSA1}). The following CSL property
computes the probability that the total LHY level remains
in the range $[0,e]$ between 96 and 500 hours.
\vspace*{-.2ex}
$$\vspace*{-.2ex}\Prob{\Globally{[96,500]} (LHY\_c + LHY\_n \leq 0 + e)}$$

%The results for different values of $e$, reported in
%Table~\ref{tab:DD:constant}, show that LHY is always smaller than
%10, while there is a small probability that it exceeds $e$ for $0
%\leq e \leq 9$, quantifying the observed trend.

The results reported in Table~\ref{tab:DD:constant} quantify the
observed trend, showing that there is a small probability that LHY
exceeds $e$ for $0 \leq e \leq 20$, and that this probability decreases with increasing $e$.

%\subsubsection{LL}

%Random graphs from PRISM

%\begin{figure}[hbtp]
%\centering
%\includegraphics[width=0.49\textwidth]{Figs/LL_time_around_average_value_133_1run.png}
%\includegraphics[width=0.49\textwidth]{Figs/LL_time_around_average_value_133_10run.png}\\
%\includegraphics[width=0.49\textwidth]{Figs/LL_time_around_average_value_133_100run.png}
%  \caption{LL probability being at medium.}\label{}
%\end{figure}

%\begin{figure}[hbtp]
%\centering
%\includegraphics[width=0.49\textwidth]{Figs/LL_time_around_average_value_160_100run.png}
%\includegraphics[width=0.49\textwidth]{Figs/LL_time_around_average_value_range_100run.png}
%  \caption{LL probability being at medium.}\label{}
%\end{figure}

%\subsubsection{LD}

%Random graphs from PRISM

%\begin{figure}[hbtp]
%\centering
%\includegraphics[width=0.49\textwidth]{Figs/P_max_at_i.png}
%\includegraphics[width=0.49\textwidth]{Figs/P_est_max_at_T.png}
%  \caption{LD probability max is at level i. prob being at estimated max at time T.}\label{}
%\end{figure}

\subsection{Distribution of Mutational Effects: robustness analysis}
\label{sec:methods_results:dme}

%Robustness to noise is an important feature of biological systems.
Thus far we have investigated aspects of the inherent random noise
caused by the small size of the system and the discrete nature of its
components. We have also explored the consequences of variations
in the light environment. We now consider the impact of mutational noise
caused by DNA changes that alter
%functional properties of the system by:
%(i) adding new reactions; (ii) effectively switching off existing
%reactions or (iii) changing
reaction rates by affecting the
structure of corresponding proteins.
%
%% Change start  2010-01-26 LLoewe
%CT as->in that
This analysis differs from the above in that it requires simulation of vast numbers of different parameter combinations.
%
%% Change end  2010-01-26 LLoewe
%
%Focusing on robustness with respect to the last type of mutational noise, we use aspects of
We do this in the context of a recently developed framework for evolutionary
systems biology that describes how the effects of DNA changes propagate
through various levels of organisation and abstraction until they impact fitness
at the highest level of biological functionality~\cite{loewe09}.
We limit our analysis to what has been described as the ``third'' level of the adaptive
landscape~\cite{loewe09}, which maps a combination of biochemical reaction
rates to a computable system-level property likely to affect fitness via a quantitative mechanistic model.
%We leave it to structural biologists to
%determine how amino acid sequence changes alter reaction rates; future
%studies will also have to investigate how fitness is affected by the system-level
%property we investigate.

\begin{figure}[b!]
\centering
  \includegraphics[width=\textwidth]{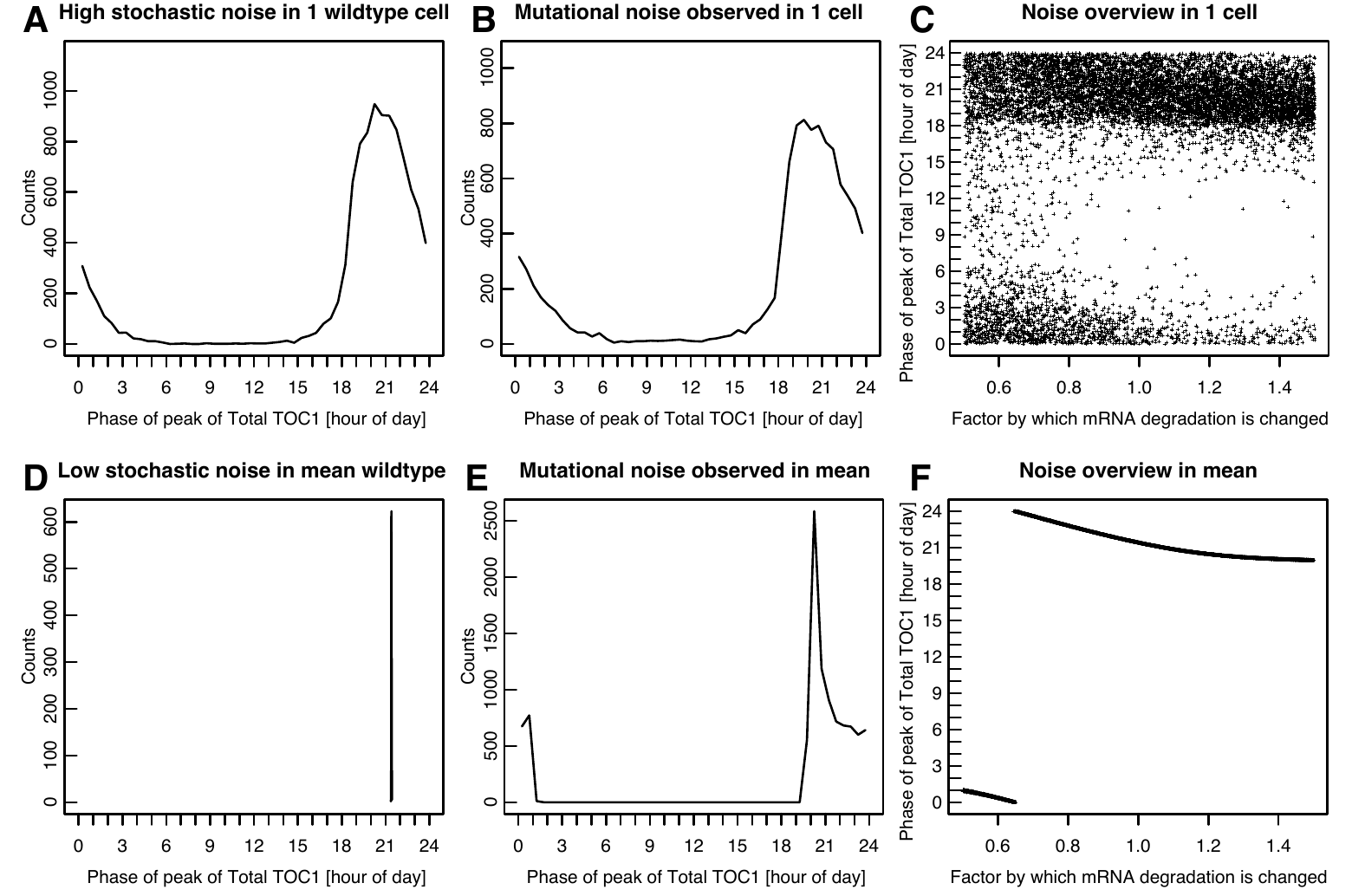}

  \caption{Robustness analysis showing how inherent stochastic
noise and mutational noise affect the peak phase of total TOC1 in 4 sets of 10000 runs. Here, noise is measured by the width of the corresponding distributions in phase values, where wider widths indicate greater noise.
Plots (a--c) are dominated by the high internal stochastic noise observed in single-cell simulations ($
\Omega = 50$). Plots (d--f) observe population averages as computed with $\Omega = 50
\times 10^6$ and show virtually no internal stochastic noise. The first column (a,d) plots the behaviour of
the wild-type, the second (b,e) adds mutational
noise by changing the mRNA degradation rate by a factor drawn from a uniform distribution [0.5,1.5], where 1.0 is the wild-type. The third column (c,f) shows how phase depends on mutational effects on mRNA degradation.
%%
%Black points
%and lines show the last single phase observation in a
%simulation and yellow the mean of 11 phase
%observations for $\Omega = 50 \times 10^6$ (at $\Omega = 50$ cycles were missed, so 10.29 $\pm$
%0.86 sd for WT and 9.63 $\pm$  1.39 for Mutant).
%% Results are from 4 sets of 10000 runs.
}\label{fig:DME_ComboPlot}
\end{figure}

We generated a StochKit~\cite{liEtAl08} model using
the Bio-PEPA Workbench~\cite{biopepa_site} in
order to build on the code base of an analysis framework that has previously
been used to investigate the evolutionary systems biology of
%
%% Change start  2010-01-26 LLoewe
a different, highly simplified circadian clock system that lacks entrainment~\cite{loewe-hillston08}. %% LL: this takes care of differentiating the old from the new paper.
We added a class for computing phase in particularly noisy circadian clocks, using  two thresholds to block stochastic noise from generating
artificially short `cycles' with very low amplitude. Thresholds were set at 20\% and 35\% of the distance between the highest peak and lowest trough
observed over 10 days, since minima are less variable than maxima (see Figure~\ref{fig:LD_all:12_12_SSA1}).
We used the peak of total TOC1 as a phase marker since our model checking results showed peaks to have higher signal/noise ratios (Figure~\ref{fig:mc_distribution_1_day}(b)).

We consider the parameter combination used in the rest of the paper to be the wild-type and
introduce mutational noise by multiplying degradation rates of all mRNAs in the system (deg7 and deg9 in
Figure~\ref{fig:model_diagram}) by a uniformly distributed factor ([0.5,1.5], where 1.0 represents the wild-type).  %% LL: just shortened.
To differentiate between internal stochastic noise and mutational noise we ran 40000 simulations in
two sets, each analysing 10000 time-courses for the wild-type and 10000 for mutants.
In the first set (Figure~\ref{fig:DME_ComboPlot}(a--c)), the system size $\Omega = 50$ corresponds to that
of a single cell. The resulting noise is substantial
as indicated by the large width of the distribution of observed phase values
(Figure~\ref{fig:DME_ComboPlot}(a)).
%
%This makes it difficult to observe
%mutational noise (Figure~
%\ref{fig:DME_ComboPlot}(a) and (b) are difficult to distinguish).
In the second set of simulations (Figure~\ref{fig:DME_ComboPlot}(d--f)), we increased  $\Omega$ a
million fold. This results in an excellent approximation of corresponding time-courses produced by ODEs; these
simulations are thus denoted as `mean' here.
%
%We assume that this is equivalent to observing the mean behaviour of a million cells,
%but that depends on various details of the system, which we do not have the space to investigate here.
%It suffices to say that this leads to time-courses which are equivalent to those from ODEs to a very good approximation.
%% LL: Since I did the comparison, I could add:
%(better than averaging over the equivalent number of single cell SSA simulations).
% We used this approximation to investigate the inherent noise that comes with observing the mean
%behaviour in experiments that typically involve about  a million cells.
The resulting internal stochastic noise
is minimal (see the extremely narrow distribution in Figure~\ref{fig:DME_ComboPlot}(d)).
%On both wild-type backgrounds we introduced mutational noise in the form of a factor that
%changes the degradation rates of all mRNAs in the system (deg7 and deg9 in Figure~\ref{fig:model_diagram}). We draw this factor
%from a uniform distribution [0.5; 1.5], where 1.0 represents the wild-type.
%

Our `mean' results show
%clearly
how phase is affected by mutational  changes in the mRNA degradation rate (Figure~\ref{fig:DME_ComboPlot}(d,e)). If the same mutational changes are introduced in the noisy single cell system, they are much more difficult to detect (see
% e.g. compare the almost identical distributions in
Figure~\ref{fig:DME_ComboPlot}(a,b)).   %% LL just cutting words.
%%As a result variability is introduced into the phase,
%but the noise of single cell observations almost completely obscures these
%changes (see plots in the middle of Figure~\ref{fig:DME_ComboPlot}).
%
As a different way of visualising results,
Figure~\ref{fig:DME_ComboPlot}(c,f) plots the high-level consequences of
mutations (phase) against their low-level effect (the factor affecting mRNA
degradation). The resulting graphs for the `mean' system show clearly how a
change in mRNA degradation rate is expected to affect the mean change in
phase (Figure~\ref{fig:DME_ComboPlot}(f)). 
Figure~\ref{fig:DME_ComboPlot}(c) shows the corresponding plot for single
cell simulations. It demonstrates that a wave of TOC1 peaks starts around
the time of dusk (18h in these simulations) and continues for a few hours.
The time at which the wave starts is minimally affected by the mutations we
investigated, in stark contrast to the variance, which is much lower for
higher mRNA degradation rates. In other words, a lower rate of mRNA
degradation leads to greater internal stochastic noise. This increase in
variance explains why the average phase is strongly shifted towards later
hours for smaller mRNA degradation rates in the `mean' system
(Figure~\ref{fig:DME_ComboPlot}(f)), since the mean is strongly affected by
large values in skewed distributions as found in (Figure~\ref{fig:DME_ComboPlot}(c)).
%
%. The large number of wild-type samples in these plots exposes the extraordinary amount of
%noise in the single cell system (see `vertical line' at $x = 1.0$ in lower left plot of Figure~\ref{fig:DME_ComboPlot}
%that is generated by the additional samples of the wild-type).
Taken together, these results demonstrate that a higher mRNA degradation rate will on average move the phase forward
and make it more reliable (decrease the phase variance), whereas a decrease will move it backwards and make it less reliable.

These subtle patterns in the mutational robustness of the clock could not have been uncovered without running large numbers of simulations with varying parameter combinations. Such work requires a different infrastructure for data analysis from projects that analyse only a few parameter sets.

%increase the interval of the day during which most peaks for TOC1 occur.

%We observed some interesting effects when computing the mean of phase
%observations within one simulation. This apparently innocuous procedure has
%non-trivial distorting effects on the distribution of phase values in noisy
%simulations (compare single phase values (black) with arithmetic means of
%$\leqq11$ phase values (yellow) in Figure~\ref{fig:DME_ComboPlot}). This is
%caused by the modular nature of phase, with small values (`early morning')
%following larger ones (`late night'), invalidating some means and inducing a
%shift to earlier values. Thus care must be taken when interpreting means.

%% LL: Text above is the result of cutting down the following:
%The nature of this distortion becomes obvious once
%`low' phase values are taken into consideration. Due to the modular nature
%of phase (`round the clock') small values (`early morning') follow after big values
%(`late night'). Computing a mean of these does not give a mean time of an event,
%but rather a strangely skewed distribution that is misleadingly shifted towards
%`earlier' values. While this is not easily corrected by shifting phase in large samples for single cells (virtually every hour of the day receives at least a few peaks), this is almost no problem for most mutants, when ODE-like means are computed (which are probably not needed in this case). Thus care must be taken when using means in computing the
%raw data that enters distributions.

\section{Conclusions}
\label{sec:conclusions}

We have studied the circadian network of \textit{Ostreococcus tauri}
by developing a process algebra model of the clock, based on an
existing deterministic representation that was parameterised according to
quantitative experimental data. We have investigated several key
aspects of the clock, such as the conditions necessary for
persistent oscillations in its constituent genes and proteins, as well as
the effect of different environmental and mutational changes on the
phases of these oscillations. We used the Bio-PEPA stochastic process
algebra as a modelling language and applied a range of the analysis
methods supported. Because of the low copy numbers of the
molecular species involved in the clock network, we focused on
stochastic analysis methods which enable the system's intrinsic
variability to be observed.

In particular, we used stochastic simulation to explore how the clock
responds to changes in the light environment, and compared the results
obtained against the behaviour of the corresponding deterministic system. We
predict that the qualitative behaviour of the free-running (LL) clock will
be dependent on the size of the cellular population; while damped
oscillations will be observed in large populations (simulated by the SSA
average and the deterministic model), self-sustained oscillations may be
detectable in single cells (simulated by individual runs of the SSA).
Model-checking was further used to investigate how the variability of the
clock's behaviour changes over a circadian cycle. By computing
the time-dependent probability distributions of the
clock proteins, we identified the time of peak expression as the most robust
phase marker, suggesting its use as an
experimental measure.

Finally, we added mutational noise to our system by
randomly changing the overall rate of mRNA degradation and observing
how this affects the phase of the oscillations of a key clock
protein, likely to have an impact on fitness. We found that the large
amount of stochastic noise at the single cell level makes it hard to
observe functional changes that may be induced by mutations, without
averaging over many observations.
%We also addressed the nontrivial issue of computing the mean of
%modular variables such as the phase of oscillations in a clock system.
%OEA: are we still addressing this issue?
%MLG: I commented out the sentence

A number of the novel hypotheses we have formulated in this
modelling study may provide new biological insights into the behaviour of
the \textit{Ostreococcus} clock and will hopefully inspire subsequent experimental
research. In addition, further theoretical work can build on the novel model-checking results
reported here, to explore additional ways in which systems biology models can be
automatically analysed using approaches based on concurrency theory. Our results
demonstrate that  the integration of different computational techniques is critical for fully
quantifying the architectural~\cite{Akman08} and mutational~\cite{loewe09} robustness
of the circadian clock.

%and improve the way entrainment is modelled in CTMC.
%

\vspace*{-0.5ex}
\subsection*{Acknowledgements}
The authors thank Gerben van Ooijen for TopCount data and
Jane Hillston and Andrew Millar for their helpful
comments. The Centre for Systems Biology at Edinburgh is a Centre
for Integrative Systems Biology (CISB) funded by BBSRC and EPSRC,
ref.\ BB/D019621/1. CT is supported by The International Human Frontier Science Program Organization.

\bibliography{biblio}
\bibliographystyle{eptcs} % or whatever you prefer

\newpage
\appendix
\section*{Appendix}
\section{The full Bio-PEPA model}
\label{sec:appendix:biopepa}

\noindent \textbf{Kinetic parameters:}
\\[1ex]
\begin{tabular}[l]{l l l | l l l}
$\mathit{acc\_rate}$ & = & $0.085759993119922787$ \quad & \quad $\mathit{R\_toc1\_lhy}$ & = & $0.80473130211377397$\\
$\mathit{H\_toc1\_lhy}$ & = & $2.4786793492076216$ \quad & \quad $\mathit{L\_toc1}$ & = & $0.0001028030683282734$\\
$\mathit{R\_toc1\_acc}$ & = & $0.40030354494924164$ \quad & \quad $\mathit{D\_mrna\_toc1}$ & = & $0.33395900070057227$\\
$\mathit{T\_toc1}$ & = & $0.65069237578254624$ \quad & \quad $\mathit{Di\_toc1\_ia\_l}$ & = & $0.11696163098006726$\\
$\mathit{Di\_toc1\_ia\_d}$ & = & $0.34434576584349563$ \quad & \quad $\mathit{D\_toc1\_a\_l}$ & = & $0.53999998111757508$\\
$\mathit{D\_toc1\_a\_d}$ & = & $0.3587344573844497$ \quad & \quad $\mathit{H\_lhy\_toc1}$ & = & $2.4123768479176113$\\
$\mathit{R\_lhy\_toc1\_a\_l}$ & = & $3.3859126401378155$ \quad & \quad $\mathit{R\_lhy\_toc1\_a\_d}$ & = & $1.1074418532202324$\\
$\mathit{D\_mrna\_lhy}$ & = & $1.9405472466939$ \quad & \quad $\mathit{T\_lhy}$ & = & $6.5204407183218498$\\
$\mathit{Di\_lhy\_cn}$ & = & $7.0630744698933485$ \quad & \quad $\mathit{D\_lhy\_l}$ & = & $0.34866585983482207$\\
$\mathit{D\_lhy\_d}$ & = & $0.21098655584281875$ \quad & & &
\end{tabular}\\[2ex]

\noindent \textbf{Time of the day at which dawn and dusk occur:}
\\[1ex]
\begin{tabular}[l]{l l l l l}
$t_{\mathit{dawn}}$ & = & $6$ & & \\
$t_{\mathit{dusk}}$ & = & $18$ & // & for the LD 12:12 system
\end{tabular}\\[2ex]

\noindent \textbf{Time-dependent function representing light in LD system:}
\\
$$\mathit{light\_time} = H\left( \left(\left(\mathit{time} - 24 \cdot \left\lfloor \frac{\mathit{time}}{24} \right\rfloor\right) - t_{\mathit{dawn}}\right) \cdot \left(t_{\mathit{dusk}} - \left(\mathit{time} - 24 \cdot \left\lfloor \frac{\mathit{time}}{24} \right\rfloor\right)\right)\right)$$

\noindent \textbf{Scaling factor:}
\\[1ex]
\begin{tabular}[l]{l l l}
$\Omega$ & = & $50$
\end{tabular}\\[1.2ex]

\noindent \textbf{Initial values:}
\\[1ex]
\begin{tabular}[l]{l l l}
$\mathit{acc\_init}$ & = & $\left\lfloor 0.99897249736755245 \cdot \Omega \right\rfloor$\\
$\mathit{TOC1\_mRNA\_init}$ & = & $\left\lfloor 1.9315264449894309e^{-06} \cdot \Omega \right\rfloor$\\
$\mathit{TOC1\_i\_init}$ & = & $\left\lfloor 0.34581773957827311 \cdot \Omega \right\rfloor$\\
$\mathit{TOC1\_a\_init}$ & = & $\left\lfloor 0.47960829226604956 \cdot \Omega \right\rfloor$\\
$\mathit{LHY\_mRNA\_init}$ & = & $\left\lfloor 9.9999999999999995e^{-07} \cdot \Omega \right\rfloor$\\
$\mathit{LHY\_c\_init}$ & = & $\left\lfloor 4.0361051173018776 \cdot \Omega \right\rfloor$\\
$\mathit{LHY\_n\_init}$ & = & $\left\lfloor 6.7029410613103796e^{-06} \cdot \Omega \right\rfloor$
\end{tabular}\\[2ex]

\noindent \textbf{Additional functions:}
\\[1ex]
\begin{tabular}[l]{l l l}
$\mathit{tmp\_toc1\_transcription}$ & = & $\mathit{L\_toc1} + \mathit{acc} \cdot \frac{\mathit{R\_toc1\_acc}}{\Omega}$\\[0.6ex]
$\mathit{toc1\_a\_decay}$ & = & $\mathit{light\_time} \cdot \mathit{D\_toc1\_a\_l} + \left(1-\mathit{light\_time}\right) \cdot \mathit{D\_toc1\_a\_d}$\\[0.3ex]
$\mathit{toc1\_i\_a\_conversion}$ & = & $\mathit{light\_time} \cdot \mathit{Di\_toc1\_ia\_l} + \left(1-\mathit{light\_time}\right) \cdot \mathit{Di\_toc1\_ia\_d}$\\[0.3ex]
$\mathit{lhy\_decay}$ & = & $\mathit{light\_time} \cdot \mathit{D\_lhy\_l} + \left(1-\mathit{light\_time}\right) \cdot \mathit{D\_lhy\_d}$\\[0.6ex]
$\mathit{lhy\_toc1\_reg}$ & = & $\mathit{TOC1\_a} \cdot \left(\mathit{light\_time} \cdot \frac{\mathit{R\_lhy\_toc1\_a\_l}}{\Omega} + \left(1-\mathit{light\_time}\right) \cdot \frac{\mathit{R\_lhy\_toc1\_a\_d}}{\Omega}\right)$
\end{tabular}\\[2ex]

\noindent \textbf{Functional Rates:}
\\[2ex]
\begin{tabular}[l]{l @{~~} l @{~~~} l l}
$\mathit{prod}_1$  & : & $\mathit{acc\_rate} \cdot \Omega \cdot \mathit{light\_time}$ & //~ light accumulator increase: mass action\\[0.3ex]
$\mathit{deg}_2$  & : & $\mathit{acc\_rate} \cdot \mathit{acc}$ & //~ light accumulator decrease: mass action\\[0.6ex]
$\mathit{transc}_3$  & : & $\Omega \cdot \frac{\mathit{tmp\_toc1\_transcription}}{1 + \mathit{tmp\_toc1\_transcription} + \left(\frac{\mathit{R\_toc1\_lhy}}{\Omega} \cdot \mathit{LHY\_n}\right)^{\mathit{H\_toc1\_lhy}}}$ & //~ TOC1 transcription: Hill kinetics\\[0.6ex]
$\mathit{deg}_4$  & : & $\mathit{toc1\_a\_decay} \cdot \mathit{TOC1\_a}$ & //~ TOC1 degradation: mass action\\[0.3ex]
$\mathit{transl}_5$  & : & $\mathit{T\_toc1} \cdot \mathit{TOC1\_mRNA}$ & //~ TOC1 translation: mass action\\[0.3ex]
$\mathit{conv}_6$  & : & $\mathit{TOC1\_i\_a\_conversion} \cdot \mathit{TOC1\_i}$ & //~ TOC1 conversion: mass action\\[0.3ex]
$\mathit{deg}_7$  & : & $\mathit{D\_mrna\_toc1} \cdot \mathit{TOC1\_mRNA}$ & //~ TOC1 mRNA degradation: mass action\\[0.6ex]
$\mathit{transc}_8$  & : & $\Omega \cdot \frac{\mathit{lhy\_toc1\_reg}^{\mathit{H\_lhy\_toc1}}}{1 + \mathit{lhy\_toc1\_reg}^{\mathit{H\_lhy\_toc1}}}$ & //~ LHY transcription: Hill kinetics\\[0.9ex]
$\mathit{deg}_9$  & : & $\mathit{D\_mrna\_lhy} \cdot \mathit{LHY\_mRNA}$ & //~ LHY mRNA degradation: mass action\\[0.3ex]
$\mathit{transl}_{10}$ & : & $\mathit{T\_lhy} \cdot \mathit{LHY\_mRNA}$ & //~ LHY translation: mass action\\[0.3ex]
$\mathit{transp}_{11}$ & : & $\mathit{Di\_lhy\_cn} \cdot \mathit{LHY\_c}$ & //~ LHY nuclear transport: mass action\\[0.3ex]
$\mathit{deg}_{12}$ & : & $\mathit{lhy\_decay} \cdot \mathit{LHY\_c}$ & //~ LHY degradation, cytosol: mass action\\[0.3ex]
$\mathit{deg}_{13}$ & : & $\mathit{lhy\_decay} \cdot \mathit{LHY\_n}$ & //~ LHY degradation, nucleus: mass action
\end{tabular}\\[4.5ex]

\noindent \textbf{Species components:}
\\[2ex]
\begin{tabular}[l]{l l l}
$\mathit{LHY\_c}$ & $\rmdef$ & $\mathit{transl}_{10} \, \product \ + \ \mathit{transp}_{11} \, \reactant \ + \ \mathit{deg}_{12} \, \reactant$\\
$\mathit{LHY\_mRNA}$ & $\rmdef$ & $\mathit{transc}_8 \, \product \ + \ \mathit{deg}_9 \, \reactant \ + \ \mathit{transl}_{10} \, \activator$\\
$\mathit{TOC1\_a}$ & $\rmdef$ & $\mathit{deg}_4 \, \reactant \ + \ \mathit{conv}_6 \, \product \ + \ \mathit{transc}_8 \, \activator$\\
$\mathit{TOC1\_mRNA}$ & $\rmdef$ & $\mathit{transc}_3 \, \product \ + \ \mathit{transl}_5 \, \activator \ + \ \mathit{deg}_7 \, \reactant$\\
$\mathit{acc}$ & $\rmdef$ & $\mathit{prod}_1 \, \product \ + \ \mathit{deg}_2 \, \reactant \ + \ \mathit{transc}_3 \, \activator$\\
$\mathit{TOC1\_i}$ & $\rmdef$ & $\mathit{transl}_5 \, \product \ + \ \mathit{conv}_6 \, \reactant$\\
$\mathit{LHY\_n}$ & $\rmdef$ & $\mathit{transc}_3 \, \inhibitor \ + \ \mathit{transp}_{11} \, \product \ + \ \mathit{deg}_{13} \, \reactant$
\end{tabular}\\[4ex]

\noindent \textbf{Species observables:}
\\[2ex]
\begin{tabular}[l]{l l l}
$\mathit{Total\_LHY}$ & = & $\mathit{LHY\_c} + \mathit{LHY\_n}$\\
$\mathit{Total\_TOC1}$ & = & $\mathit{TOC1\_i} + \mathit{TOC1\_a}$
\end{tabular}\\[3.5ex]

\noindent \textbf{Model component:}
\\[1ex]
$$\mathit{LHY\_c}(\mathit{LHY\_c\_init}) \ \sync{*} \ \mathit{LHY\_mRNA}(\mathit{LHY\_mRNA\_init}) \ \sync{*} \ \mathit{TOC1\_a}(\mathit{TOC1\_a\_init}) \ \sync{*}$$\vspace*{-3ex}
$$\mathit{TOC1\_mRNA}(\mathit{TOC1\_mRNA_init}) \ \sync{*} \ \mathit{acc}(\mathit{acc\_init}) \ \sync{*} \ \mathit{TOC1\_i}(\mathit{TOC1\_i\_init}) \ \sync{*} \ \mathit{LHY\_n}(\mathit{LHY\_n\_init})$$\vspace*{5ex}

\section{Modelling light/dark cycles in PRISM}
\label{sec:appendix:prism}

\begin{scriptsize}
\begin{verbatim}
const int max_days_simulated = 21;

module time

 min : [0..59] init 0;
 hour : [0..23] init 0;
 day : [0..max_days_simulated] init 0;
 light_time : [0..1] init 0;

 [change_min] (min < 59) -> 60: (min'=min + 1);

 [change_hour_dawn] (min = 59 & hour = (tdawn-1) ) -> 60: (min'=0) & (hour' = hour + 1) & (light_time' = 1);
 [change_hour_dusk] (min = 59 & hour = (tdusk-1) ) -> 60: (min'=0) & (hour' = hour + 1) & (light_time' = 0);
 [change_hour] (min = 59 & hour < 23 & hour != (tdawn-1) & (hour != tdusk-1) ) ->
                                                            60: (min'=0) & (hour' = hour + 1);

 [time_change_day] (min = 59 & hour = 23 & day < max_days_simulated) ->
                                                            60: (min'=0) & (hour'=0) & (day' = day + 1);
 [time_end_day] (min = 59 & hour = 23 & day = max_days_simulated) -> 60: (min'=0) & (hour'=0) & (day'=0);

endmodule
\end{verbatim}
\end{scriptsize}

\section{Additional simulation and analysis results}
\label{sec:appendix:results}

\begin{figure}[hbtp]
\centering
\subfigure[LL -- ODE]{\includegraphics[width=0.32\textwidth]{Figs/LL_ODE.pdf}\label{fig:omega:LL_ODE}}
\subfigure[LL -- SSA $\Omega=50$]{\includegraphics[width=0.32\textwidth]{Figs/LL_SSA_10000run.pdf}\label{fig:omega:LL_SSA10000_50}}
\subfigure[LL -- SSA $\Omega=500$]{\includegraphics[width=0.32\textwidth]{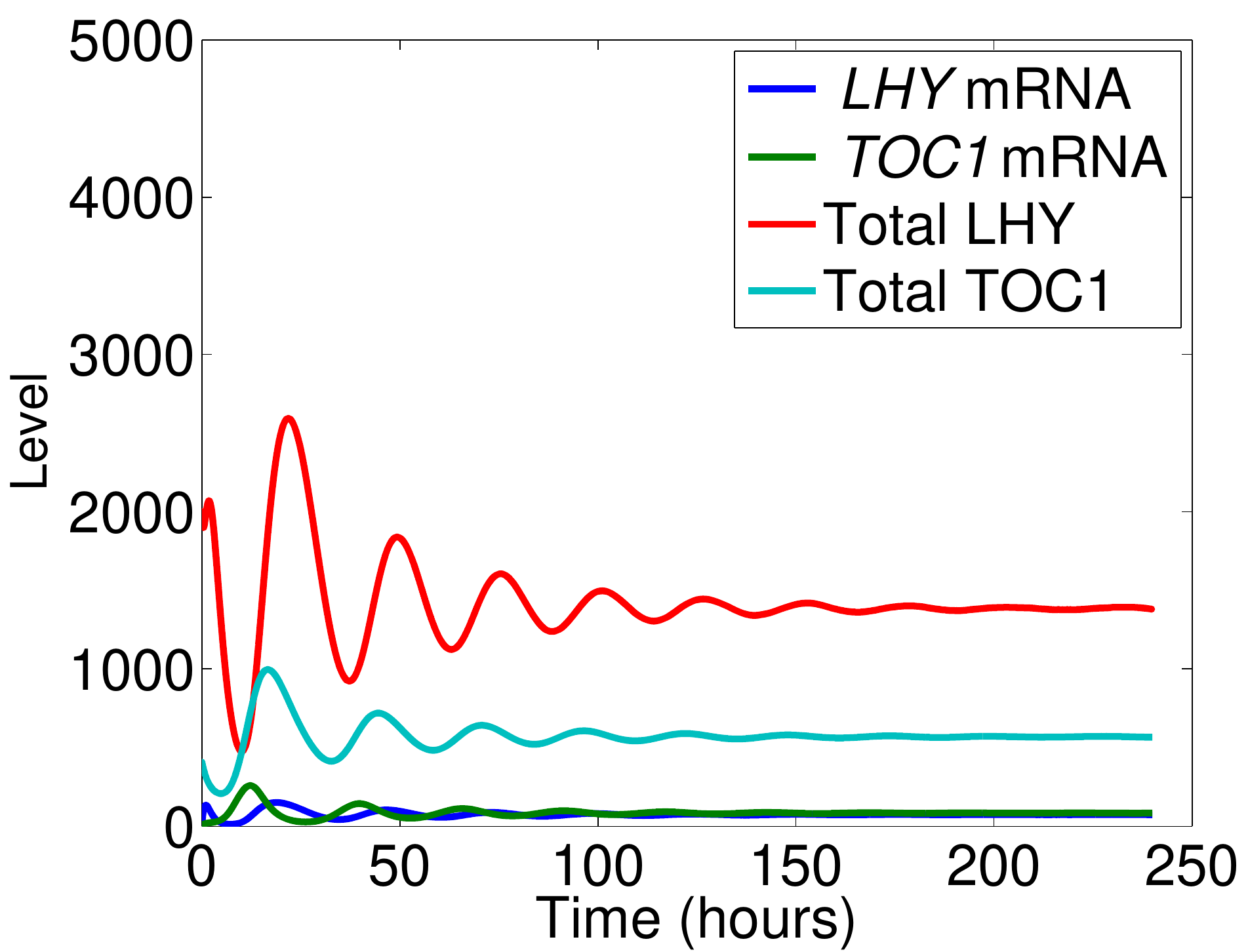}\label{fig:omega:LL_SSA10000_500}}
  \caption{Comparison of deterministic and stochastic models for constant light (LL) with different scaling factors $\Omega$.}\label{fig:omega}
\end{figure}

Figure~\ref{fig:omega} shows the comparison between the ODE results
and the mean SSA behaviour with scaling factors $\Omega=50$ and
$\Omega=500$. We observe a slight difference between the SSA results
with the different scaling factors. The reference value is $\Omega=50$
(estimated from experimental protein counts), and consequently the predicted
biological behaviour is that shown in
Figure~\ref{fig:omega:LL_SSA10000_50}. Note, instead, that the ODE
results (Figure~\ref{fig:omega:LL_ODE}) agree perfectly with the SSA
results for the larger value $\Omega=500$
(Figure~\ref{fig:omega:LL_SSA10000_500}).

\begin{table}[hbt]
\centering
\begin{small}
\begin{tabular}{|c|c|} \hline
DD & LL \\ \hline \hline
$-0.0858$ & $-0.0858$ \\ \hline
$-0.2110$ & $-0.0200-0.2509i$ \\ \hline
$-0.3340$ & $-0.0200+0.2509i$ \\ \hline
$-0.3443$ & $-0.6447-0.3216i$ \\ \hline
$-0.3587$ & $-0.6447+0.3216i$ \\ \hline
$-1.9405$ & $-1.9506$ \\ \hline
$-7.274$1 & $-7.4117$ \\ \hline
\end{tabular}
\end{small}
\caption{Eigenvalues of the linearised ODE system about the DD and
LL fixed points.}\label{tab:evalues:LLDD}
\end{table}

Table~\ref{tab:evalues:LLDD} lists the eigenvalues of linearisation at the fixed points of the deterministic model. These determine the dynamics in the vicinity of the steady-states~\cite{Guckenheimer83}. All eigenvalues of the DD fixed point are negative and real, identifying it as a stable node~\cite{Guckenheimer83}. The LL fixed point
retains 3 of the DD eigenvalues;  the remaining ones comprise two complex conjugate pairs with negative real parts. The steady-state is therefore a stable focus~\cite{Guckenheimer83}. The positions of the fixed points were estimated using the Nelder-Mead simplex algorithm~\cite{Lagarias98}, as implemented in the MATLAB routine \texttt{fminsearch}. Derivatives were computed analytically using the MATLAB Symbolic Math Toolbox.

\end{document}